\documentclass[10pt, conference, letterpaper]{IEEEtran}

\IEEEoverridecommandlockouts

\ifCLASSINFOpdf
\else
\fi
  \usepackage{tipa}
  \usepackage{graphicx}
	\usepackage{epsfig}
	\usepackage{epstopdf}
	\usepackage{upgreek}
	\usepackage[nolist,nohyperlinks]{acronym}

\usepackage[cmex10]{amsmath}
\usepackage{subfigure}

\usepackage{cite}

\begin{acronym}
\acro{NoC}{Network-on-Chip}
\acro{WNoC}{Wireless Network-on-Chip}
\acro{EM}{Electromagnetic}
\acro{SiO$_2$}{Silicon Dioxide}
\acro{AIN}{Aluminum nitride}
\acro{SiP}{System-in-Package}
\acro{SDM}{Software-defined metamaterial}
\acro{PDP}{Power Delay Profile}
\acro{mm-wave}{millimeter-wave}
\acro{MAC}{Medium Access Control}
\acro{OOK}{On-Off Keying}
\acro{ISI}{Inter-Symbol Interference}
\acro{TSV}{Through-Silicon Via}
\acro{CSMA}{Carrier Sense Multiple Access}
\acro{BER}{Bit Error Rate}
\acro{RZ}{Return-to-Zero}
\end{acronym}

\usepackage{listings}
\begin{document}



\title{Engineer the Channel and Adapt to it: \\Enabling Wireless Intra-Chip Communication}


\author{
Xavier Timoneda, Sergi Abadal, Antonio Franques, Dionysios Manessis, Jin Zhou,\\
Josep Torrellas, Eduard Alarc\'{o}n, Albert Cabellos-Aparicio
\thanks{The authors gratefully acknowledge support from the Spanish MINECO under grant PCIN-2015-012, from the EU's H2020 FET-OPEN program under grants No. 736876 and No. 863337, and by the Catalan Institution for Research and Advanced Studies (ICREA).}
\thanks{Xavier Timoneda, Sergi Abadal, Eduard Alarc\'{o}n and Albert Cabellos-Aparicio are with NaNoNetworking Center in Catalunya (N3Cat) at Universitat Polit\`{e}cnica de Catalunya (UPC), Barcelona, Spain (e-mails: xavier.timoneda@upc.edu, abadal@ac.upc.edu, eduard.alarcon@upc.edu, acabello@ac.upc.edu)}
\thanks{Antonio Franques and Josep Torrellas are with the Department of Computer Science at University of Illinois at Urbana-Champaign (UIUC), Illinois, USA (e-mails: franque2@illinois.edu, torrella@illinois.edu)}
\thanks{Jin Zhou is with the Department of Electrical and Computer Engineering at University of Illinois at Urbana-Champaign (UIUC), Illinois, USA (e-mail: jinzhou@illinois.edu)}
\thanks{Dionysios Manessis is with System Integration \& Interconnection Technologies at Fraunhofer Institute for Reliability and Microintegration (IZM), Berlin, Germany (e-mail: dionysios.manessis@izm.fraunhofer.de)}

}


%


\maketitle


\begin{abstract}
Ubiquitous multicore processors nowadays rely on an integrated packet-switched network for cores to exchange and share data. The performance of these intra-chip networks is a key determinant of the processor speed and, at high core counts, becomes an important bottleneck due to scalability issues. To address this, several works propose the use of mm-wave wireless interconnects for intra-chip communication and demonstrate that, thanks to their low-latency broadcast and system-level flexibility, this new paradigm could break the scalability barriers of current multicore architectures. However, these same works assume 10+ Gb/s speeds and efficiencies close to 1 pJ/bit without a proper understanding on the wireless intra-chip channel. This paper first demonstrates that such assumptions do not hold in the context of commercial chips by evaluating losses and dispersion in them. Then, we leverage the system's monolithic nature to \emph{engineer the channel}, this is, to optimize its frequency response by carefully choosing the chip package dimensions. Finally, we exploit the static nature of the channel to \emph{adapt to it}, pushing efficiency-speed limits with simple tweaks at the physical layer. Our methods reduce the path loss and delay spread of a simulated commercial chip by 47 dB and 7.3$\times$, respectively, enabling intra-chip wireless communications over 10 Gb/s and only 3.1 dB away from the dispersion-free case.
\end{abstract}



%
\IEEEpeerreviewmaketitle


\acresetall


	
	
	


\section{Introduction} \label{sec:introduction} 
Multicore processors are present in virtually every computing domain nowadays. They integrate a number of processor cores within the same chip and, in the past few years, manufacturers have been consistently increasing the core count seeking higher execution speeds. However, in order to translate this potential into effective performance, the on-chip communication problem must be solved: cores need an integrated interconnect to exchange or share data and, for densely populated chips, traditional interconnects are burdensome and slow down the processor. Communication, not computation, thus becomes the main performance bottleneck in multicore systems \cite{Marculescu2009}. 

\begin{figure}[!t]
\centering
\includegraphics[width=\columnwidth]{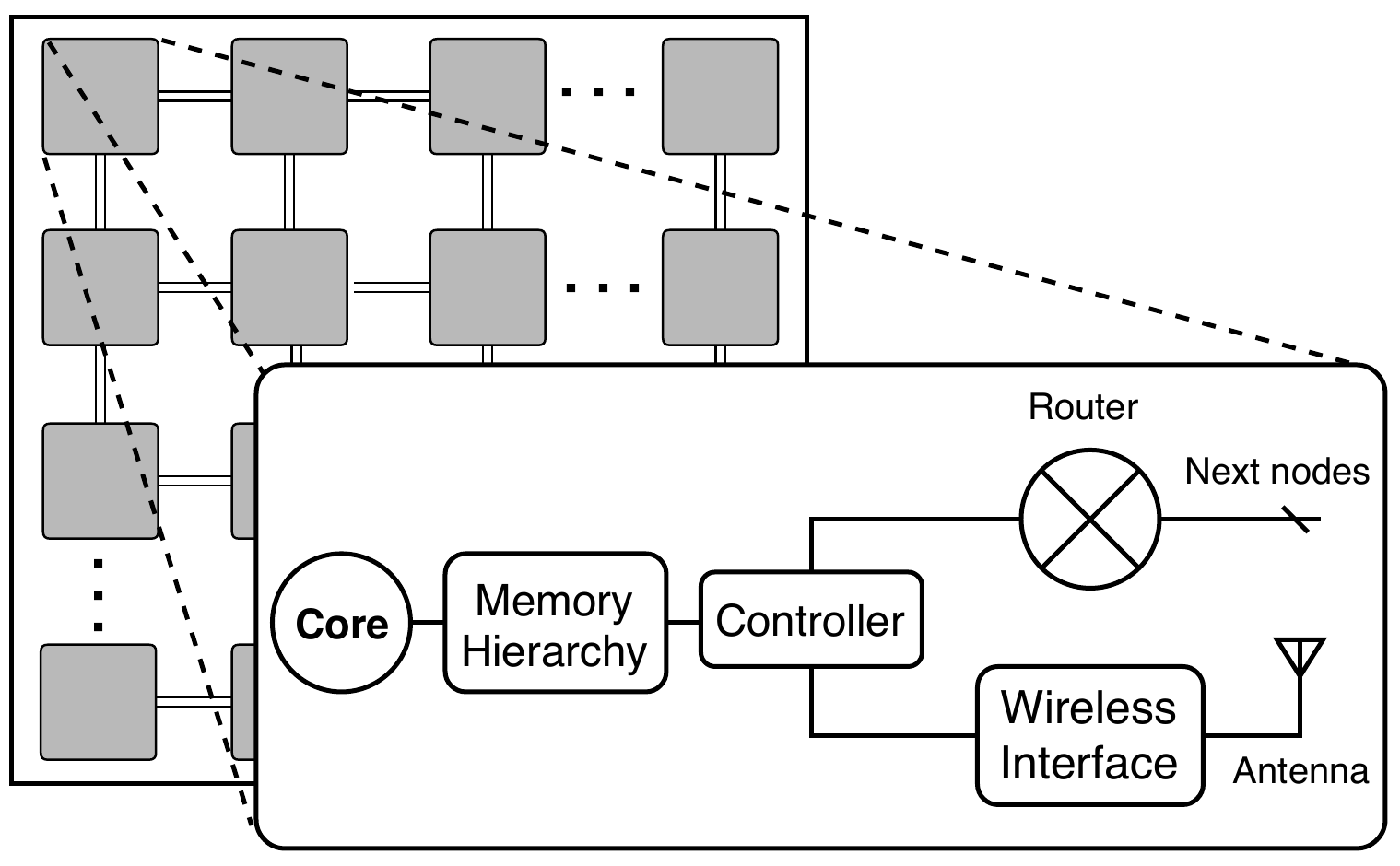}
\vspace{-0.5cm}
\caption{Sketch of a Wireless Network-on-Chip architecture.}
\label{fig:scenario}
\vspace{-0.2cm}
\end{figure}

In the past, most chips did not contain more than a handful of cores and on-chip communication was easily performed through a bus. Since buses do not scale well with the number of cores, a completely different approach was soon required. The adopted solution, called \ac{NoC}, consists of a packet-switched network of routers that are co-integrated with the cores as represented in Figure \ref{fig:scenario}. Since then, \acp{NoC} have been widely applied not only in research works \cite{Vangal2008, Nychis2012, Park2012a, Chen2015a}, but also in commercial chips such as Tilera's TILE-GX \cite{Wentzlaff2007} or Intel's Xeon Phi \cite{XeonPhi}. Nevertheless, with the arrival of extreme scaling and massive multicore architectures, standard \acp{NoC} start to show performance and efficiency issues~\cite{Bertozzi2014}. New paradigms are thus required in the manycore era.

The scalability problems of \acp{NoC} are mainly the network diameter and overprovisioning. As further elaborated in Sec. \ref{sec:background}, these cause the communication latency and power to increase, especially for chip-wide transactions. Therefore, any new candidate to improve existing \acp{NoC} should address them and, among a few alternatives \cite{Kim2012Survey}, \ac{WNoC} shows great promise in this regard. In short, \ac{WNoC} basically consists in overlaying a set of wireless intra-chip links over a backbone wired \ac{NoC}. This reduces the latency of chip-wide transfers, including broadcasts, by virtue of the omnidirectional speed-of-light propagation of radio waves, and also combats overprovisioning thanks to its global reconfigurability \cite{Matolak2012}. As shown in the literature, these unique features become key enablers of new multicore architectures capable of pushing current scalability limits \cite{AbadalMICRO, Kim2016, Sikder2016}.

The \ac{WNoC} paradigm builds on the foundations of widespread \ac{mm-wave} technology. A wide variety of on-chip antennas is already available \cite{Markish2015, Cheema2013, Wu2017b} and wireless intra-chip communication with such antennas has been experimentally confirmed in multiple works \cite{Floyd2002, Zhang2007, Wu2013a}. Additionally, 60/90 GHz integrated transceivers specifically designed for \ac{WNoC} have been tested \cite{Yu2014, Laha2015, AbadalTON, Mineo2015}. On top of this, a great variety of works have evaluated new topologies and routing protocols \cite{Abadal2018, Sujay2012, Gade2017a, DiTomaso2015, Choi2018, Abadal2018a} in an attempt to exploit the potential of \ac{WNoC} at the network level. 

The main caveat of the majority of \ac{WNoC} research is that it lays on incorrect channel models. Many works \cite{abadal2019wave, Yan2009, Zhang2007, Chen2009, Yeh2013, Narde2019, Gade2017, Rayess2017, Elmasri2019} either neglect the influence of the chip package, which introduce losses and dispersion, or directly neglect dispersion whatsoever. This does not invalidate the potential of the \ac{WNoC} paradigm, but leads to erroneous assumptions on the achievable speed and power. For instance, many \ac{WNoC} architectures assume bandwidths well over 10 GHz~\cite{Kim2016, DiTomaso2015, Choi2018, AbadalASPLOS, Fernando2019}, which may not be achievable due to multipath effects. Other works obtain power consumption estimates by assuming path losses between below 30 dB \cite{Yu2014a, Yu2015, Subramaniam2017, Ahmed2018}. In the present study, we show that these assumptions are false for standard chip packages.

This paper aims to fill this gap and restate the potential of \ac{WNoC} by proposing, as the main contribution, a novel co-design methodology that (i) properly characterizes the wireless intra-chip channel, and (ii) identifies and exploits its uniquenesses. It can be summarized in three pillars:
\begin{itemize}
\item \textbf{Channel characterization:} we study the propagation within a realistic computing package, which has been often overlooked. Frequency and time domain analyses are performed to extract attenuation and dispersion scaling trends. With this, we prove that the assumptions made in most \ac{WNoC} works may not hold true, and that path loss and delay spread often follow contradicting trends.
\item \textbf{Channel engineering:} the intra-chip channel is unique in that it can be engineered. Therefore, we propose an optimization methodology that explores the package design space to jointly minimize attenuation (path loss) and dispersion (delay spread). We illustrate the methodology by applying it in a particular chip package design and reduce the path loss and delay spread by 30 dB and 3.52$\times$ together, respectively, or by 47 dB and 7.32$\times$ in separated extreme cases.
\item \textbf{Static transceiver optimization:} the intra-chip channel is also unique in that it is quasi-deterministic. Based on this, we propose to combat dispersion by predicting the multipath effects and adapting the transceiver back-end to them. We easily accommodate 10 Gb/s and reach beyond the coherence bandwidth limit, figures that would be unattainable with conventional coding.
\end{itemize}



Although the static and monolithic nature of the \ac{WNoC} scenario were already discussed in \cite{Matolak2013CHANNEL, Abadal2018}, this is the first work that, to the best of the authors' knowledge, systematically exploits the unique traits of the wireless intra-chip channel. The proposed methodology could potentially lead to the conditions to operate at 10--20 Gb/s with 1--2 pJ/bit, figures that are widely assumed in the literature but that would be otherwise unattainable. It is worth noting that very few other wireless communication scenarios, if any, allow to \emph{engineer the channel} to enhance propagation. 



The remainder of this paper is organized as follows. Sec. \ref{sec:background} provides some background. Sec. \ref{sec:systemmodel} details the proposed methodology, which is then evaluated in Sec. \ref{sec:simulationresults}. Finally, Sec. \ref{sec:discussion} discusses the results and Sec. \ref{sec:conclusion} concludes the paper.


\section{Background} 
\label{sec:background} 

\begin{figure*}[!t]
\centering
\includegraphics[width=\textwidth]{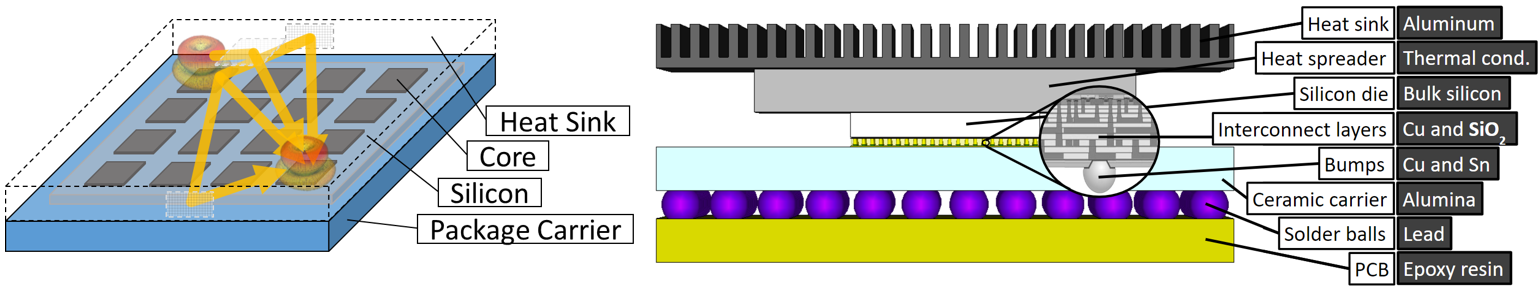}
\vspace{-0.5cm}
\caption{Wireless propagation within a computer package (left) and typical cross-section of a flip-chip package (right).}
\vspace{-0.2cm}
\label{fig:background}
\end{figure*}

\textbf{Network-on-Chip:} NoCs generally implement a 2-D mesh topology wherein every router is connected to a core and to its four neighbors (Fig. \ref{fig:background}). The choice is driven by the regularity of the topology and the short path lengths, which simplifies the routers and the links. Topologies requiring long links are in fact discouraged as their energy and delay scale exponentially with length and technology \cite{ITRS}. Short links, however, come at the cost of a network diameter that scales as $2(k-1)$ in a $k\times k$ mesh. Thus, 64-core chips, which are commercially available \cite{Wentzlaff2007, XeonPhi}, have a network diameter of 14 hops with a chip-wide latency of several tens of nanoseconds without contention. This delay would be incurred by transmissions among far-apart cores or, even worse, broadcasts that would also increase contention as they flood the mesh. Alternatively, carefully designed \acp{WNoC} can reduce this delay to a few nanoseconds regardless of the location of data and number of destinations. This difference in performance is crucial because communications are often on the critical path of the program and any added delay can slow down execution \cite{AbadalMICRO}. 

\textbf{Wireless Network-on-Chip:} \ac{WNoC} broadly refers to the implementation of wireless intra-chip links on top of a wired \ac{NoC}.
A packet arriving to a wireless interface is serialized, modulated and radiated by the antenna with a given pattern as we show in Figure \ref{fig:background}. Radio waves propagate through the package at nearly the speed of light until reaching the intended destinations, also located within the same package, where they are demodulated and deserialized. Since intermediate router hops are avoided, \ac{WNoC} reduces the latency of long-range and broadcast communications by an order of magnitude. On the downside, wireless bandwidth is limited and needs to be shared among the cores. 

The physical layer of \ac{WNoC} adapts to chip resource constraints. The use of \ac{mm-wave} bands allows antennas to be commensurate with cores, whereas simple modulations such as \ac{OOK} are adopted to avoid bulky or power-hungry components at the transceiver. With such low order modulations, high symbol rates are needed to reach the 10+ Gb/s speeds expected for \ac{WNoC}. This, together with the stringent \ac{BER} requirements of the scenario (10\textsuperscript{-15} to be comparable to that of a wire), makes signals particularly vulnerable to \ac{ISI}. Fortunately, multipath effects can be mitigated through package--transceiver co-design, as we propose in this work. 

At the MAC and network layers, WNoCs are constrained by the resource limitations and latency requirements of the scenario. Directive antennas are prohibitive, with few exceptions, rendering spatial multiplexing impractical. Thus, MAC protocols generally rely on some variant of low-latency collision-avoiding token passing to share one or a few frequency channels, thus leaving ISI as the main source of interference \cite{Sujay2012,Abadal2018a}. Moreover, by design, wireless intra-chip networks are generally one-hop in an attempt satisfy the strong latency demands of multiprocessors \cite{Kim2016, Gade2017a, AbadalASPLOS, Fernando2019}. As a result, to support the WNoC functionalities, a routing algorithm only needs add the logic to decide when a packet should enter the wireless plane, i.e., for broadcast or long-range transmissions \cite{Abadal2018}.
The simplicity at the network layer makes the wireless option more scalable than other emerging technologies, as briefly discussed in Section \ref{sec:discussion}.

 
\textbf{Chip Structure and Antenna Placement:} 
The typical cross-section of a standard chip consists of a metal stack with 5--10 layers, separated by an insulator and placed over a lossy silicon substrate \cite{Markish2015}. Chips are then generally covered by a package that provides mechanical support and facilitates its interfacing with the rest of components. Flip-chip packages, wherein the chip is flipped over and connected to the PCB board through solder bumps, are currently widespread and preferred over wired bonding. As shown in Figure \ref{fig:background}, the chip ends up surrounded by (i) a metallic heat sink contacted by a heat spreader and (ii) the package carrier, with several metal layers on top the PCB. 

The flip-chip package does not leave much space for the antennas. Due to the presence of solder bumps, antennas cannot be implemented in the first metal layer anymore \cite{Branch2005}. Alternatively, designers have to use the metal layers closer to the silicon or, as proposed recently, drill \ac{TSV} to implement vertical monopoles \cite{Pano2019, Timoneda2018}. Due to the very stringent area constraints of the scenario, directional antennas and MIMO arrays are generally prohibitive, with few exceptions \cite{Mondal2016}.

\textbf{Chip-scale Channel Characterization:} At the chip scale, most channel characterization works have been based on full-wave simulation due to manufacturing costs and the complexity of probing in highly integrated packages \cite{Narde2019, Wang2016AWPL, Gade2017}. In open packages, however, experimental works have been more common and have shown a reasonable agreement between measurement and simulation \cite{Kim2001, Branch2005, Narde2019}. Several of those works described two propagation aspects worth considering. First, the low resistivity silicon used to facilitate transistor operation introduces significant losses and, therefore, shall be avoided \cite{Markish2015}. Second, materials used as heat spreader like Aluminum Nitride (AIN) introduce low electrical losses and, thus, would enhance propagation \cite{Yan2009}. This opens interesting perspectives to the manufacturer, which can now take chip design decisions based on the potential for wireless intra-chip communication.

Being enclosed in a metallic package, electromagnetic propagation is confined within the limits of the package. Such field confinement has positive implications on security as eavesdropping or jamming are physically avoided, but also leads to strong multipath effects. This has been formulated by Matolak \emph{et al.} through micro-reverberation theory \cite{Matolak2013CHANNEL}, yet without detailing the package structure. In fact, very few studies include the chip package in their simulations or measurements and, those that do it, are limited to low frequencies or lack proper justifications on the antenna type and placement \cite{Kim2001, Branch2005, Narde2019}. Others simply assume free space over the insulator layer \cite{Narde2019, Gade2017, Zhang2007, Yeh2013, Chen2009}. 

To find analogous results, we need to refer to works at the data center cabinet scale \cite{Khademi2015}, or at the motherboard scale in desktops or laptops \cite{Chiang2010, Wu2013a, Zajic2018}, which have structural resemblances. However, the results are not directly applicable to the chip scale due to substantial differences in dimensions, materials, and antenna placement restrictions. 

Remind that, without proper understanding of the wireless channel within the package, the impact of the wireless chip-scale paradigm cannot be really assessed. In the next section, we propose a methodology to bridge this gap.



\section{System Design} 
\label{sec:systemmodel}
Our methodology provides a way to systematically co-design the chip package and the transceiver exploiting the static and monolithic nature of the system. This way, the methodology (i) validates the \ac{WNoC} concept, (ii) increases the achievable data rate, and (iii) reduces the power consumed by the transceiver circuitry. Here, we first overview our proposal and then detail its design.

\subsection{System Overview} 
\begin{figure}[!t]
\centering
\vspace{-0.2cm}
\includegraphics[width=\columnwidth]{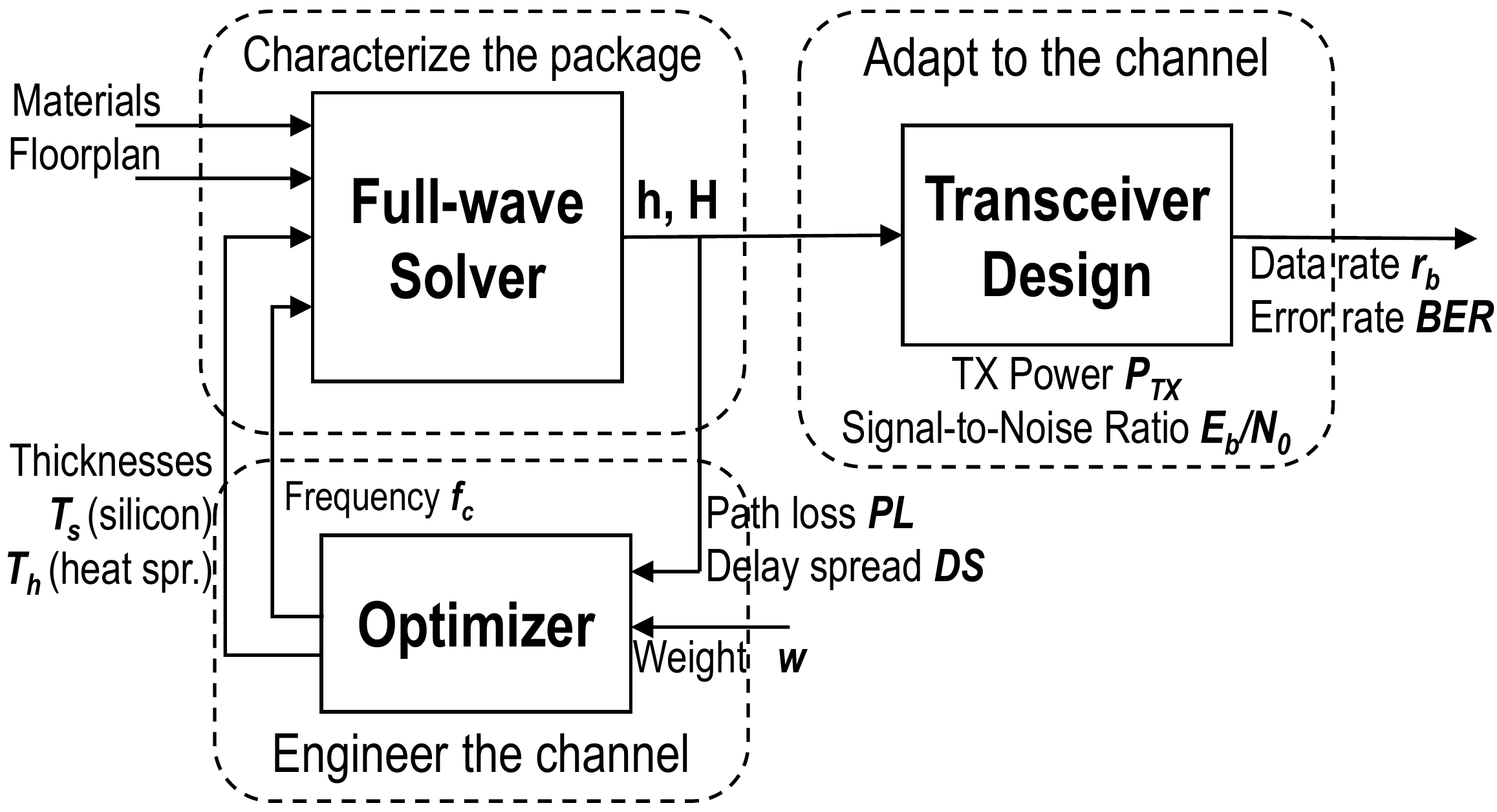}
\vspace{-0.6cm}
\caption{Proposed methodology: characterize the package, engineer the channel, and adapt to it.}
\vspace{-0.4cm}
\label{fig:system}
\end{figure}

The wireless intra-chip channel is largely unknown and prevents architects from assessing the true potential of \ac{WNoC}. The proposed methodology, summarized in Figure \ref{fig:system}, solves the problem in three steps.

First, a comprehensive characterization of the wireless channel within a chip package is performed. Through modeling and full-wave solving, we obtain the response of the wireless channel as a function of the position of the transmitting and receiving antennas within a 4$\times$4 grid, as well as two parameters that chip makers can modify at design time: the frequency band and the dimensions of the package. As further elaborated in Section \ref{sec:simulationmodel}, the results are processed to evaluate path loss and dispersion over the transmission distance.

The next step in the methodology is referred to as \emph{channel engineering} and is uniquely suited to this monolithic system. Its main goal is to find the combination of package dimensions and frequency band that jointly minimizes path loss and dispersion. To this end, we define a figure of merit that takes both aspects into account with adjustable weights, allowing manufacturers to model the importance of power and performance in the system. This figure of merit drives an optimizer that, thanks to heuristics derived from the previous characterization process, navigates through the package design tradeoffs efficiently. The exploration is possible thanks to the use of full-wave electromagnetic simulations, which avoid the need for building multiple expensive test vehicles. More details on the methodology are given in Section \ref{sec:engineering}.

Once we have found the best package and frequency band for our purposes, we optimize the transceiver by leveraging the static nature of the channel. As shown in Section \ref{sec:transceiver}, simple but effective modifications are carried out at both sides of the communication: the transmitter uses \ac{RZ} to mitigate the \ac{ISI} level, whereas the receiver uses a small and fixed set of decision thresholds to decode the current symbol based on previous bits. Both modifications are static and allow pushing the data rate beyond the typical limits imposed by the \ac{ISI} (i.e., the coherence bandwidth).

\subsection{Channel Characterization}
\label{sec:simulationmodel}

\textbf{Simulation Setup.} 
The structure shown in Figure \ref{fig:background} is modeled in CST Microwave Studio \cite{CST}, a full-wave solver. The silicon die has a resistivity of 10 $\Upomega\cdot$cm, with $\varepsilon_{r,Si} = 11.9$\; whereas the heat spreader is AIN with $\varepsilon_{r,AIN} = 9$ and negligible losses. To reduce the computational burden, the the interconnect layers and micro-bump array are approximated as a solid metallic element. This assumption has been validated in previous simulation works \cite{Timoneda2018, Timoneda2018b} and is justified by the small pitch of the interconnect layers ($<$10 $\upmu$m) and bump array ($<$0.1 mm) as compared to the excitation wavelength ($\sim$1 mm). The antenna used for the simulations is a broadband omnidirectional aperture, which allows to focus the study on the channel effects. Unless noted, we consider a homogeneous distribution of 4$\times$4 antennas within a 20$\times$20 mm\textsuperscript{2} chip and a central frequency of 60 GHz. The minimum distance among antennas at 60 GHz is 3.57$\lambda$, where the $\lambda$ is the wavelength within silicon. This distance and the high loss of silicon guarantee that there is no near-field coupling among neighboring antennas.


\textbf{Frequency Domain Analysis.} The full-wave solver uses the Finite Elements Method (FEM) to obtain the field distribution, the antenna gain, and the coupling between antennas in the frequency domain. Then, the channel frequency response $H_{ij}(f)$ is evaluated for each antenna pair as 
\begin{equation} \label{eq:Hf}
G_{i} G_{j} |H_{ij}(f)|^{2} = \frac{|S_{ji}(f)|^{2}}{(1 - |S_{ii}(f)|^{2})\cdot(1 - |S_{jj}(f)|^{2})},	
\end{equation}
where $G_{i}$ and $G_{j}$ are the transmitter and receiver antenna gains, $S_{ji}$ is the coupling between transmitter $i$ and receiver $j$, whereas $S_{ii}$ and $S_{jj}$ are the reflection coefficients at both ends \cite{Lin2007}. Once the whole matrix of frequency responses $\mathbf{H}$ is obtained, a path loss analysis can be performed by fitting the attenuation $L$ over distance $d$ to
\begin{equation} \label{eq:PL}
L = 10\,n \cdot \log_{10}(d/d_{0}) + L_{0},
\end{equation}
where $L_{0}$ is the path loss at the reference distance $d_{0}$ and $n$ is the path loss exponent \cite{Zhang2007}. The path loss exponent is around 2 in free space, below 2 in guided or enclosed structures, and above 2 in lossy environments. Since losses at the channel are crucial to determine the power consumption at the transceiver (see Section \ref{sec:discussion}) we will report improvements in terms of worst-case $L_{max}$, average $L_{avg}$, and path loss exponent $n$. 

\textbf{Time Domain Analysis.} In the time domain, we define an input excitation $x_{i}(t)$ at the input of the transmitting antenna $i$. Then, CST employs the Finite-Difference Time-Domain (FDTD) method to calculate the output signal $y_{j}(t)$ at the receiving antenna $j$. Hence, the impulse response $h_{ij}(t)$ between transmitter $i$ and receiver $j$ can be derived with the classical formulation
\begin{equation} \label{eq:ht}
y_{j}(t) = x_{i}(t) \star h_{ij}(t),
\end{equation}
where $\star$ denotes the convolution operator. Once calculated, it is straightforward to evaluate the \ac{PDP} in the channel between transmitter $i$ and receiver $j$ as
\begin{equation} \label{eq:PDP}
P_{ij}(\tau) = |h_{ij}(t,\tau)|^{2},
\end{equation}
therefore obtaining a matrix of \ac{PDP} functions $\mathbf{P}$ for all transmitters and receivers within the chip. To characterize the multipath richness of the channel, we obtain the delay spread $\tau_{rms}$ using the \ac{PDP} of each channel as 
\begin{equation}
\label{eq:DelaySpread}
\tau_{rms}^{(i,j)} = \sqrt{\frac{\int{(\tau - \overline{\tau_{ij}})^{2} P_{ij}(\tau)\, d\tau}}{\int{P_{ij}(\tau)\, d\tau}}},
\end{equation}
where $\overline{\tau_{ij}} = \frac{\int{\tau P_{ij}(\tau) d\tau}}{\int{P_{ij}(\tau)\, d\tau}}$ is the mean delay of the channel.

In this work we will assume that all wireless channels are broadcast and, therefore, they should be operated at the lowest speed ensuring correct decoding at all nodes. As a result, we will take the worst delay spread across all pairs of transmitters-receivers (i.e., across all distances) as limiting case and use it to evaluate the coherence bandwidth $B_{c}$, as follows
\begin{equation}
\label{eq:Bc}
\tau_{rms} = \max_{i,j\neq i}{\tau_{rms}^{(i,j)}} \Rightarrow B_{c} \propto \frac{1}{\tau_{rms}}.
\end{equation}
For simplicity, we will take $B_{c} = \tfrac{1}{\tau_{rms}}$.

\subsection{Channel Engineering}
\label{sec:engineering}
Our methodology takes path loss and delay spread as two metrics to be optimized. Since both aspects are dependent on multiple inputs, the channel engineering can be formally treated as a Multi-Objective Optimization (MOO) problem. These problems can be solved using algorithms amenable to MOO, such as evolutionary algorithms \cite{nam2000multiobjective}. 

Another way to tackle the channel engineering is by reducing it to a single-objective problem using weights. In particular, our methodology defines a single custom figure of merit $\phi_{w}$ that we will attempt to maximize. Since the aim is to mitigate the path loss and the delay spread, the figure of merit takes the form
\begin{equation}
\label{eq:18} 
\phi_{w} = \frac{1}{PL^{w} DS^{(1-w)}}
\end{equation}
where $PL$ is the path loss metric, $DS$ is the delay spread metric, and $w \in [0, 1]$ models the importance of power or speed in different designs. In other words, $w$ is fixed by the architect: small values will be used in high performance devices where speed needs to be optimized over power, whereas large values imply minimization of the path loss oriented to low-power embedded systems. In this paper, our metrics are $PL = L_{avg}$ and $DS = \tau_{rms}$, with $L_{avg}$ defined as the average path loss across all distances and $\tau_{rms}$ as defined in Equation \eqref{eq:Bc}. Moreover, we normalize both metrics so that they have the same dynamic range between 0 and 1.

The package engineering process as defined in this work considers three variables that can be modified at design time: the silicon thickness $T_{s}$, the heat spreader thickness $T_{h}$, and the carrier frequency $f_{c}$. Then, the objective is to maximize the figure of merit
\begin{equation}
\label{eq:opt} 
\max_{T_{s}, T_{h}, f_{c}}{\phi_{w}}\,\,,
\end{equation}
this is, to find the $T_{s}$, $T_{h}$, and $f_{c}$ values that maximize the figure of merit for a given $w$ and within the bounds given by the manufacturer or the architect\footnote{Although this work considers three key parameters, the optimization can be extended to other decisions such as antenna placement, lateral chip dimensions, or additional material choices. For instance, the resistivity of silicon could be considered as an additional optimization knob as it largely determines the losses within silicon} \cite{Zhang2007}. We conservatively assume $T_{s} \in [0.1, 0.7]$ mm and $T_{h} \in [0, 0.8]$ mm, which are ranges easily achievable with current silicon thinning and packaging techniques for 3D ICs \cite{Bieck2010}.

\begin{figure*}[!t]
\centering
\vspace{-0.2cm}
\includegraphics[width=\textwidth]{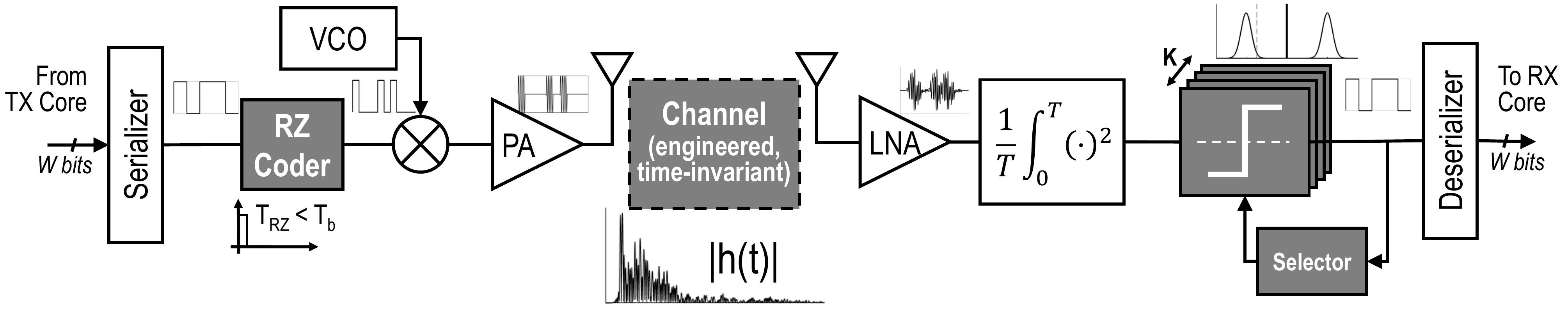}
\vspace{-0.2cm}
\caption{Physical layer in a wireless intra-chip link with \ac{OOK} and non-coherent detection. Shaded blocks identify the improvements proposed in this work.}
\vspace{-0.4cm} 
\label{fig:link}
\end{figure*} 

To solve the optimization problem, it is first worth noting that the full-wave simulations required to obtain $\phi_{w}$ for each $\{T_{s}, T_{h}, f_{c}\}$ combination are very computationally intensive, especially as $f_{c}$ increases, which renders exhaustive searching impractical. Also, path loss and dispersion are related to $\{T_{s}, T_{h}, f_{c}\}$ in non-monotonic ways and often showing opposed trends. This creates local peaks in the $\phi_{w}$ function, thus discarding methods such as the gradient-based \emph{hill climbing}, which tends to get stuck into local maxima. 

Among the pool of optimization techniques, one alternative amenable to this problem would be Simulated Annealing (SA), which uses a probabilistic method to avoid local peaks and progressively approach a global optimum. Although SA can be modified to solve MOOs \cite{nam2000multiobjective}, we treat our problem as a single-objective optimization and use conventional SA variants. Since SA has been used in other electromagnetic problems \cite{Simkin1992, Shu2004} and is widely known, we will not detail its implementation for the sake of brevity. We just note that the results of the channel characterization described in this work can help deriving the appropriate heuristics (e.g., candidate generation, cooling schedule) for SA to converge fast to the global optima. 
 



\subsection{Static Transceiver Optimization}
\label{sec:transceiver}
Once the channel is engineered to minimize path loss and delay spread, we leverage the static nature of the channel to perform simple yet effective optimizations in the RF back-end. The idea is to 
push the symbol rates while resorting to the known, deterministic channel response to keep complexity at a minimum.

Figure \ref{fig:link} shows the block diagram of a typical wireless intra-chip link. As pointed out in Section \ref{sec:background}, \ac{OOK} modulation is generally considered. Assuming a bit-energy of $E_{b} = P_{rx} / r_{b}$, where $P_{rx}$ is the received power and $r_{b}$ is the symbol rate, the \ac{BER} of \ac{OOK} is lower bounded by
\begin{equation}
\label{eq:BEROOK}
BER_{OOK} = \frac{1}{2} \mbox{erfc}\left(\sqrt{\frac{E_{b}}{4N_{0}}}\right)
\end{equation}
where \emph{erfc} is the complementary error function and $E_{b}/N_{0}$ is the signal-to-noise ratio. This bound assumes coherent detection with optimal threshold calculation and no \ac{ISI}. In our case, however, \ac{ISI} manifests when pushing the data rate beyond the Nyquist rate. To mitigate its effects, we propose two techniques: threshold adaptation and \ac{RZ} modulation.

\textbf{Threshold adaptation:} The main issue in conventional wireless environments is that multipath effects are space- and time-dependent. Therefore, its impact on the Euclidean distance between the \ac{OOK} symbols and on the optimal decision threshold cannot be predicted. In the worst case, \ac{ISI} is modeled as added noise, reducing the noise margin and leading to an approximate \ac{BER} of
\begin{equation}
\label{eq:BEROOK2}
BER_{OOK}^{isi} \approx \frac{1}{2} \mbox{erfc}\left(\sqrt{\frac{E_{b}}{4(N_{0}+I)}}\right) > BER_{OOK}
\end{equation}
where $I$ is the interference energy.

In \ac{WNoC}, the channel is time-invariant and we can calculate the exact position of each symbol at all times. This means that we can find the Euclidean distance between symbols and the optimal decision threshold for any combination of previous symbols even in the presence of \ac{ISI}. This information can be used to design a receiver composed by $K$ parallel deciders, each with its own threshold, and a register that selects the appropriate leg. Assuming that with $K$ deciders we address all \ac{ISI} effects, we can approximate the \ac{BER} as
\begin{equation}
\label{eq:BEROOK3}
BER_{OOK}^{adap} \approx \frac{1}{K} \sum_{k=1}^{K} { \frac{1}{2} \mbox{erfc}\left(\sqrt{\frac{\alpha_{k}E_{b}}{4N_{0}}}\right) }
\end{equation}
where $\alpha_{k}$ models the effect of a given past symbol combination to the Euclidean distance between current symbols. The number of required deciders scales as $K \sim \tau_{rms}/T_{b}$ where $T_{b} = 1/r_{b}$ is the symbol period assuming a binary modulation. In any case, the associated overheads are small compared to the cost of the RF front-end.

\textbf{Return to zero:} a classical way to mitigate ISI effects is by using \ac{RZ} techniques, which reduce the length of the symbol through duty cycling. One the one hand, this shortens the length of the current symbol as seen by the receiver, which implies lower spillage into the next symbols. On the other hand, the lower \ac{ISI} comes at the cost of a drop in the received energy, which may offset the gains of reduced \ac{ISI} if \ac{RZ} is not designed properly. However, since the channel is time-invariant, we can infer the duty cycle that maximizes the signal-to-interference ratio and, thus, minimizes the \ac{BER} for any symbol combination. In Equation \eqref{eq:BEROOK3}, this would be equivalent to increasing $\alpha_{k}$ for all $k$.

\section{Evaluation} 
\label{sec:simulationresults}
The three pillars of the proposed methodology are evaluated separately. Section \ref{sec:charact_eval} discusses channel scaling trends, Section \ref{sec:evalEng} shows the gains of the channel engineering process, and Section \ref{sec:evalDesign} illustrates the transceiver improvements.

\subsection{Channel Characterization} \label{sec:charact_eval}
Here, we quantify the impact of the silicon thickness $T_{s}$, the heat spreader thickness $T_{h}$ and the central frequency $f_{c}$ on the path loss and delay spread. Unless noted, we assume a homogeneous distribution of 4$\times$4 antennas and take $f_{c} = 60$ GHz and the dimensions of a standard chip ($T_{s} = 0.7$ mm and $T_{h} = 0.2$ mm) as default values. We obtain the path loss and delay spread for all antenna pairs and perform a linear regression to obtain the dependence with distance.


\begin{figure*}[!t]
\centering
\subfigure[\label{fig:siliconvsdistF}]{\includegraphics[width=0.33\textwidth]{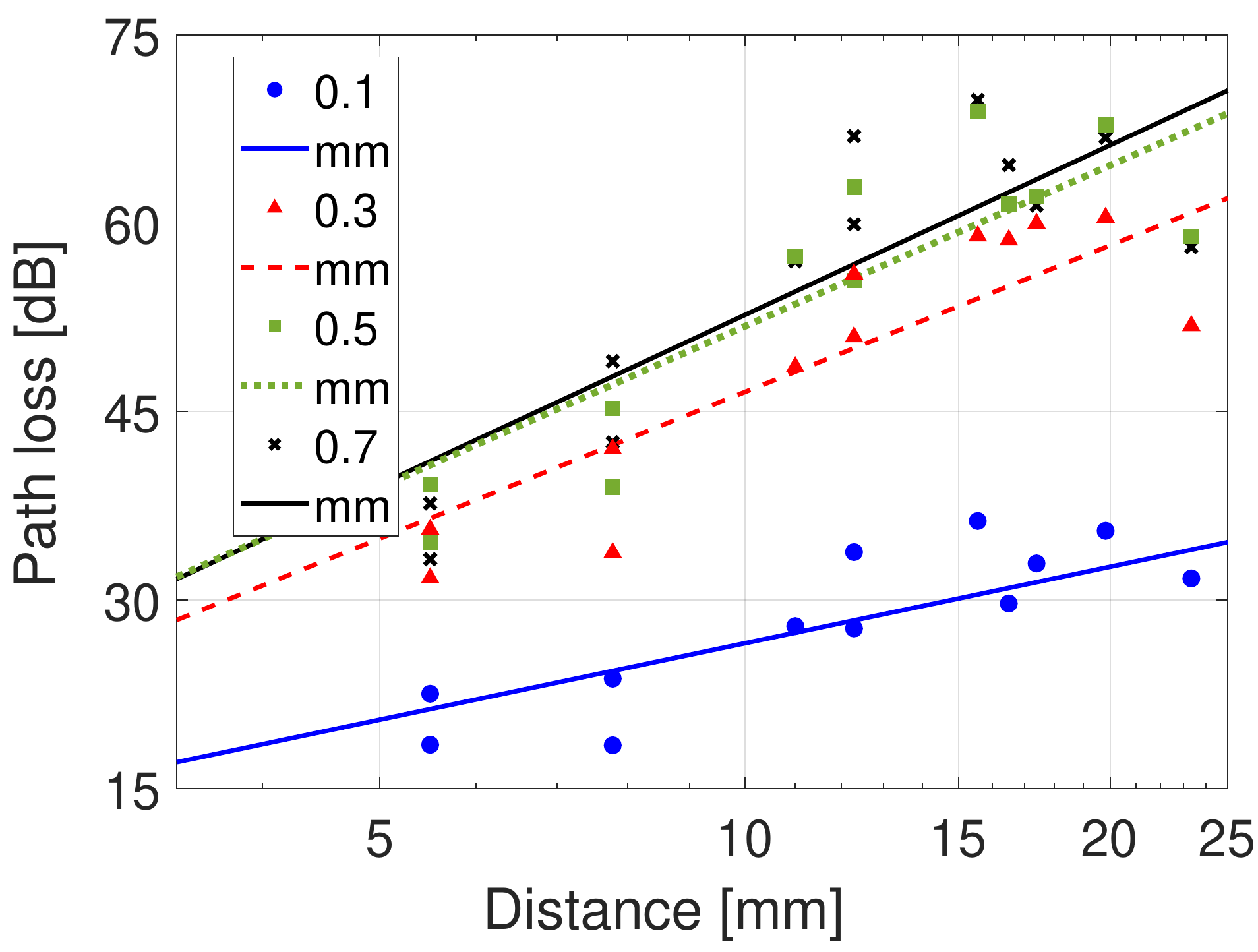}}
\subfigure[\label{fig:siliconvsdistT}]{\includegraphics[width=0.33\textwidth]{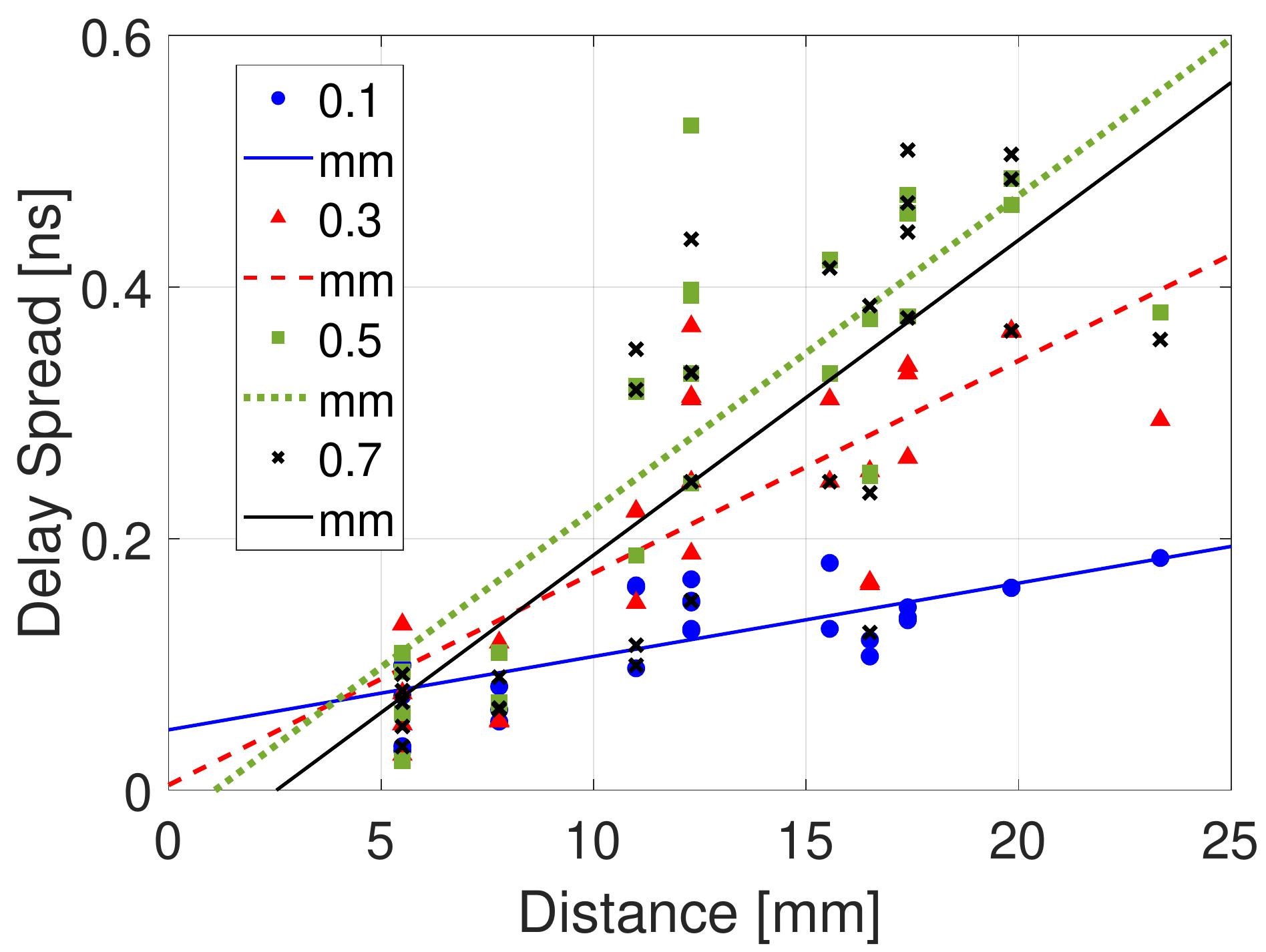}}
\subfigure[\label{fig:silicon}]{\includegraphics[width=0.3\textwidth]{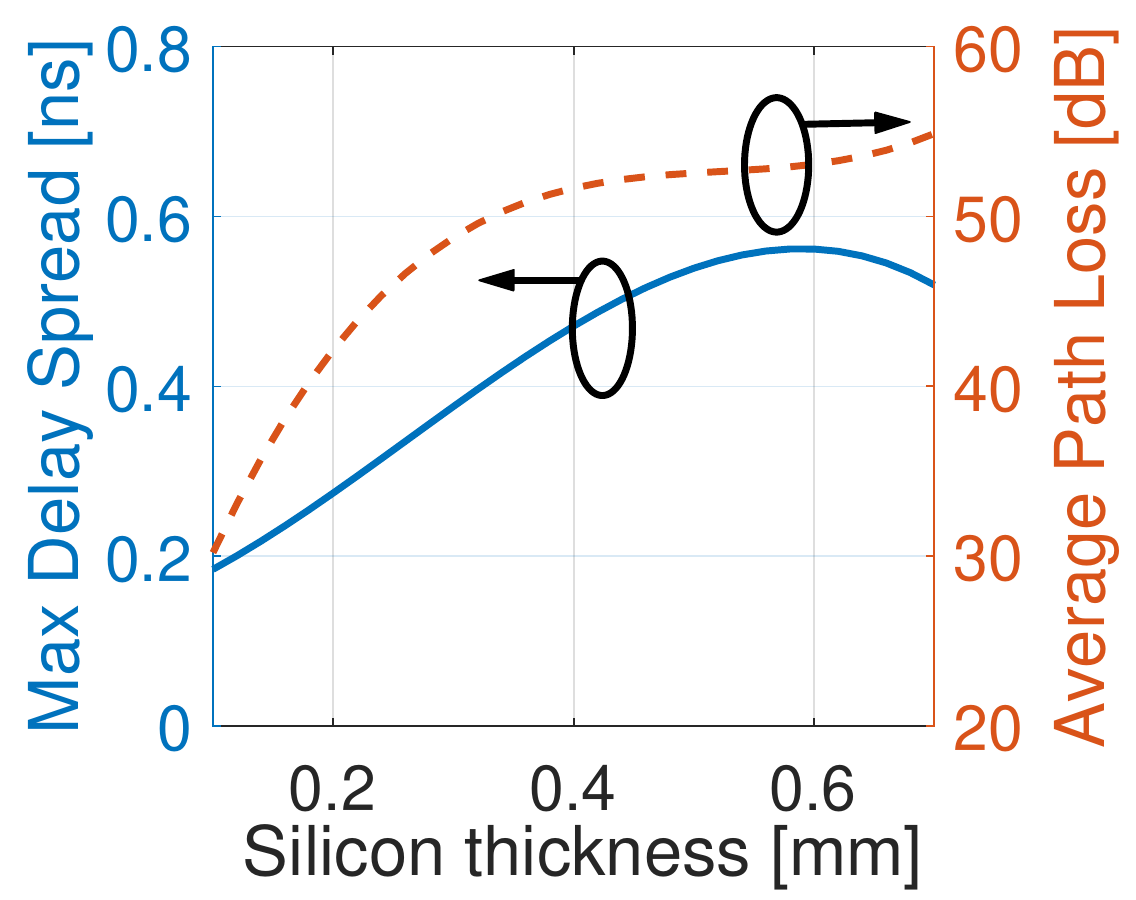}}
\vspace{-0.2cm}
\caption{Scaling analysis of the wireless channel with respect to the silicon thickness $T_{s}$ for $T_{h} = 0.2$ mm and $f = 60$ GHz, detailing (a) path loss over distance, (b) delay spread over distance, and (c) maximum delay spread $\tau_{rms}$ and average path loss $L_{avg}$ as functions of $T_{s}$.\label{fig:silicons}}
\vspace{-0.3cm}
\end{figure*} 


\begin{figure*}[!t]
\centering
\subfigure[\label{fig:AINdistF}]{\includegraphics[width=0.32\textwidth]{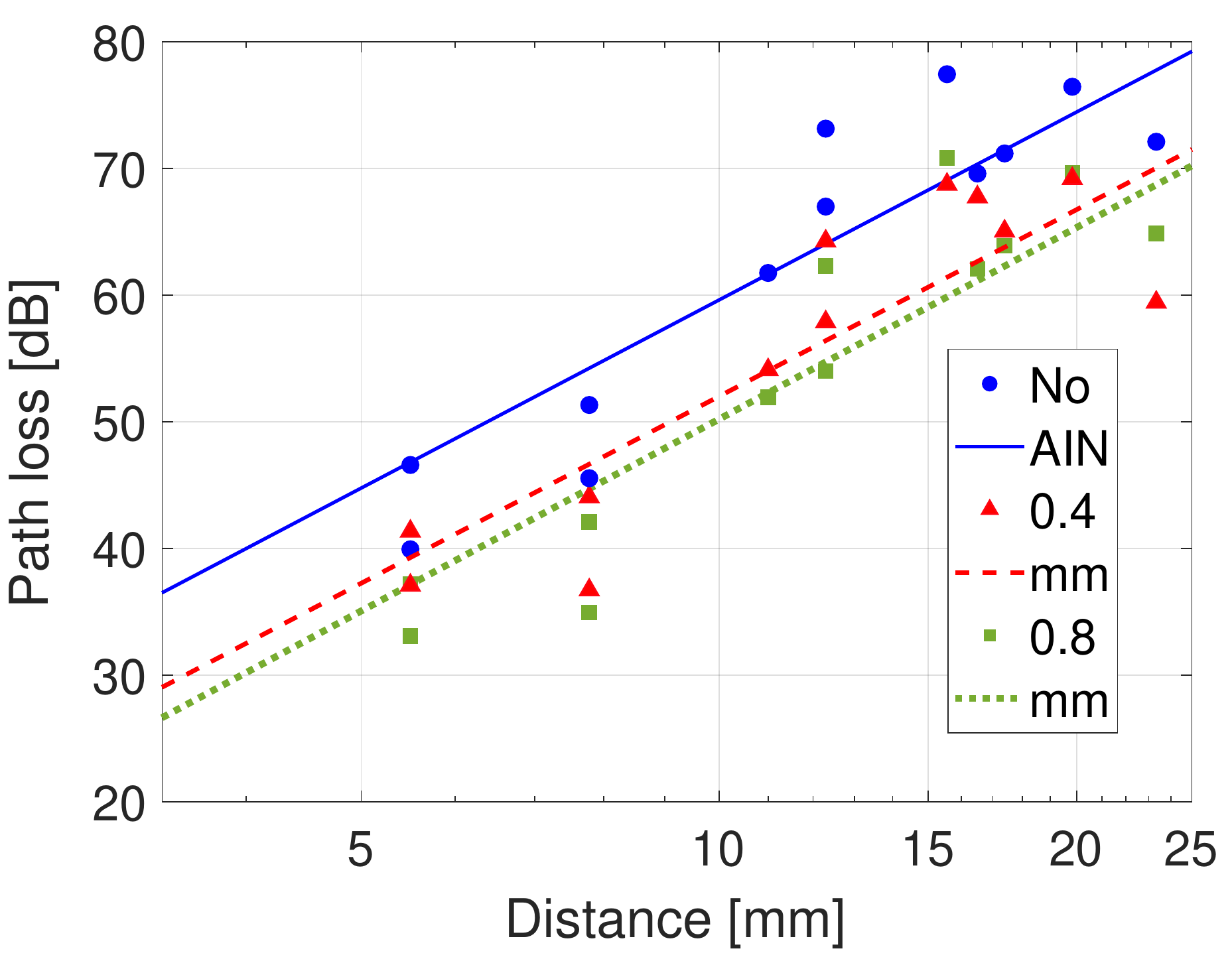}}
\subfigure[\label{fig:AINdistT}]{\includegraphics[width=0.34\textwidth]{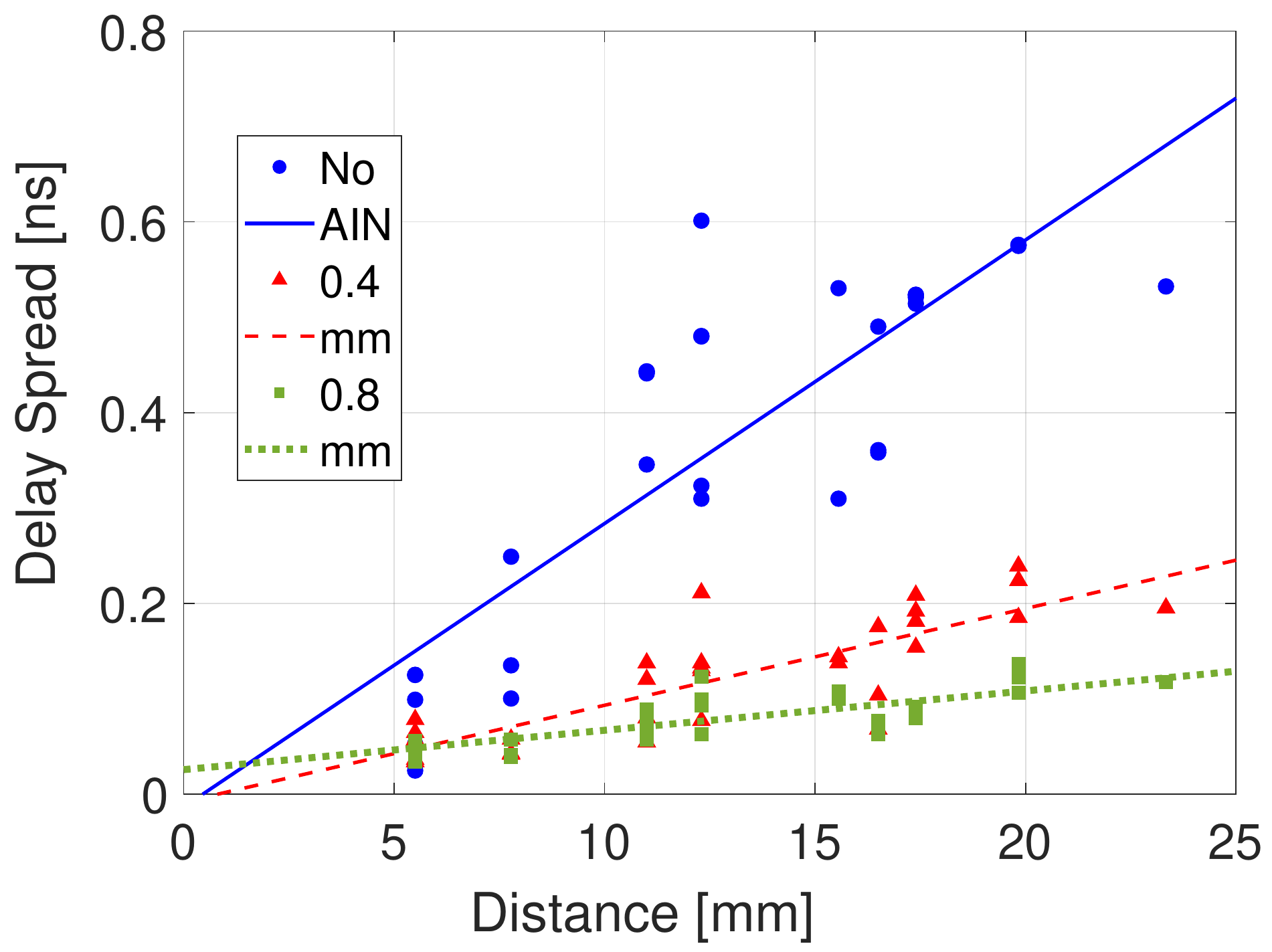}}
\subfigure[\label{fig:spreader}]{\includegraphics[width=0.3\textwidth]{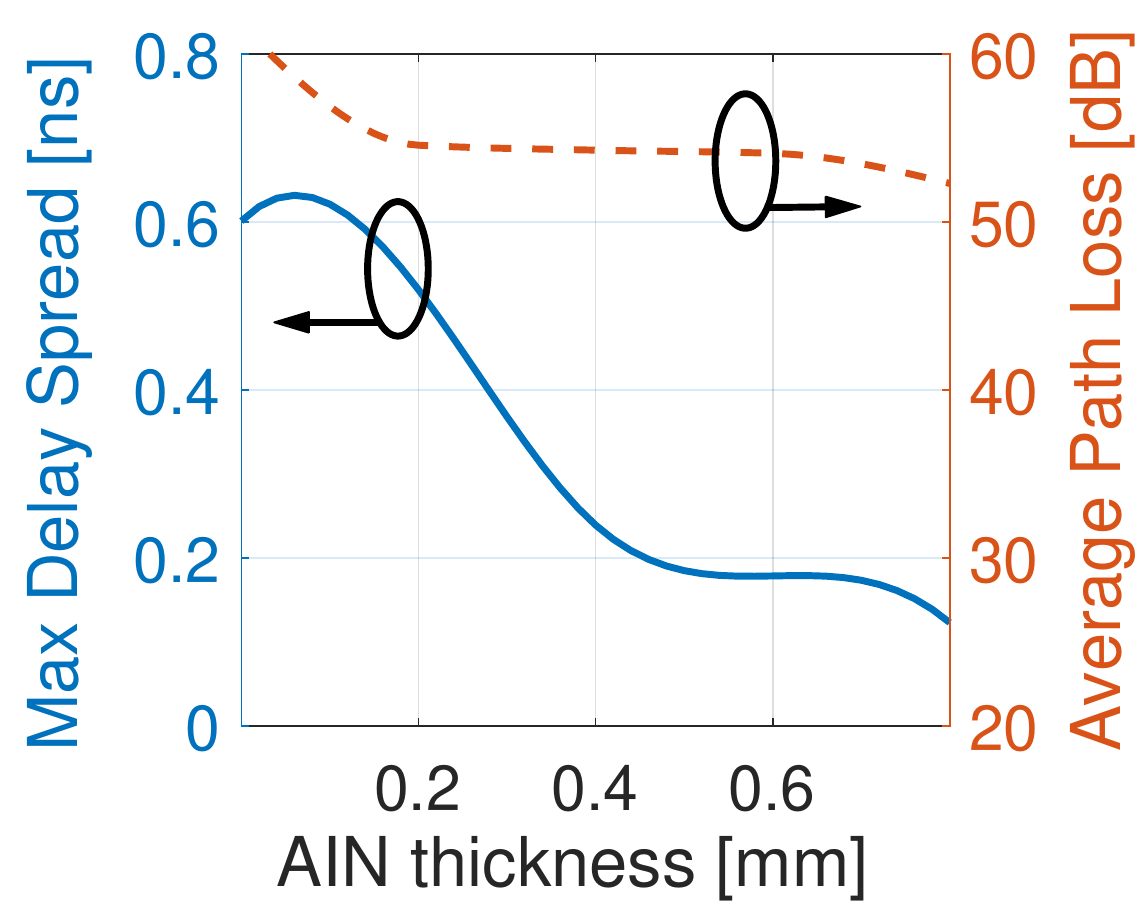}}
\vspace{-0.2cm}
\caption{Scaling analysis of the wireless channel with respect to the heat spreader thickness $T_{h}$ for $T_{s} = 0.7$ mm and $f = 60$ GHz, detailing (a) path loss over distance, (b) delay spread over distance, and (c) maximum delay spread $\tau_{rms}$ and average path loss $L_{avg}$ as functions of $T_{h}$.\label{fig:spreaders}}
\vspace{-0.3cm}
\end{figure*}


Figure \ref{fig:silicons} shows the scaling trends with respect to the silicon thickness. This layer is highly lossy, as mentioned in Sec. \ref{sec:background}, and we observe that the benefits of thinning it down are significant. A 100-$\upmu$m chip has a maximum path loss of $L_{max} = 36.29$ dB and a maximum delay spread of $\tau_{rms} = 0.19$ ns. Compared to a standard chip, the thinned alternative is 2.1$\times$ better in terms of path loss (39 dB difference) and 2.73$\times$ better in terms of worst-case delay spread (0.33 ns difference). Additionally, the path loss exponent is reduced from $n = 4.32$ to $n = 1.32$, confirming the transition from a lossy environment ($n > 2$) to a guided medium ($n < 2$). The performance also scales better in terms of delay spread, reducing the slope from 25.05 to 5.83 ps/mm.

Figure \ref{fig:spreaders} repeats the analysis by varying the heat spreader thickness $T_{h}$. Given its low electrical losses, this layer can aid propagation and its inclusion is thereby highly recommended. The delay spread improves up to 3$\times$ (from 0.6 to 0.2 ns) due to the presence of a stronger reflection cluster coming from the heat spreader. As for the path loss, the case here presented shows a limited impact in terms of path loss ($\sim$10 dB improvement in average) because most of the energy is dissipated in the 0.7-mm silicon layer before reaching the heat spreader. Although not shown due to space constraints, the effect of AIN on path loss is much more evident for thinned down silicon as the exponent drops from $n=4.01$ (no AIN) to close to $1.1$ (0.8 mm). In that case, the delay spread also oscillates between 0.2 and 0.6 ns, sometimes contradicting the path loss tendency.

%
%

\begin{figure*}[!t]
\centering
\subfigure[\label{fig:FreqdistF}]{\includegraphics[width=0.32\textwidth]{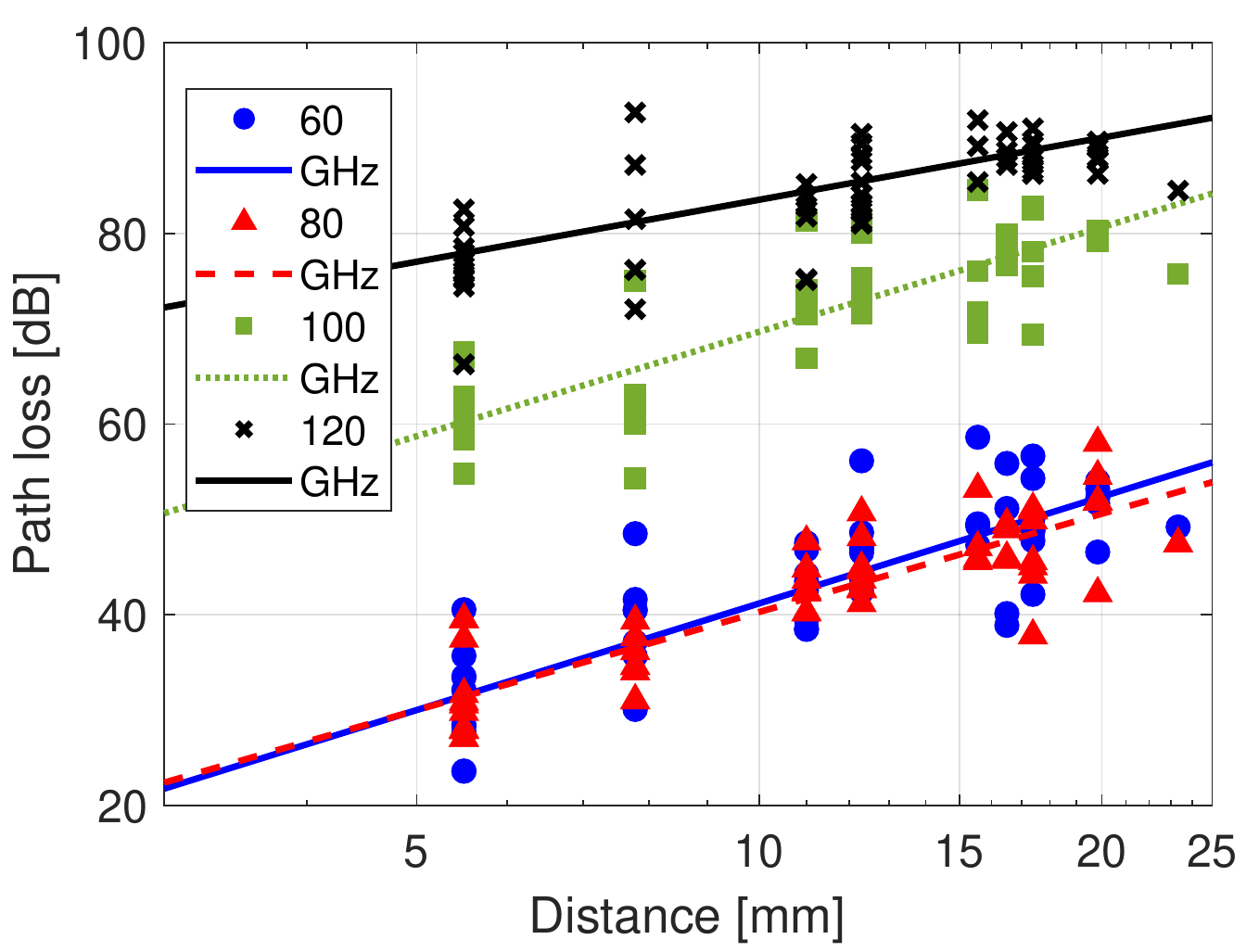}}
\subfigure[\label{fig:FreqdistT}]{\includegraphics[width=0.32\textwidth]{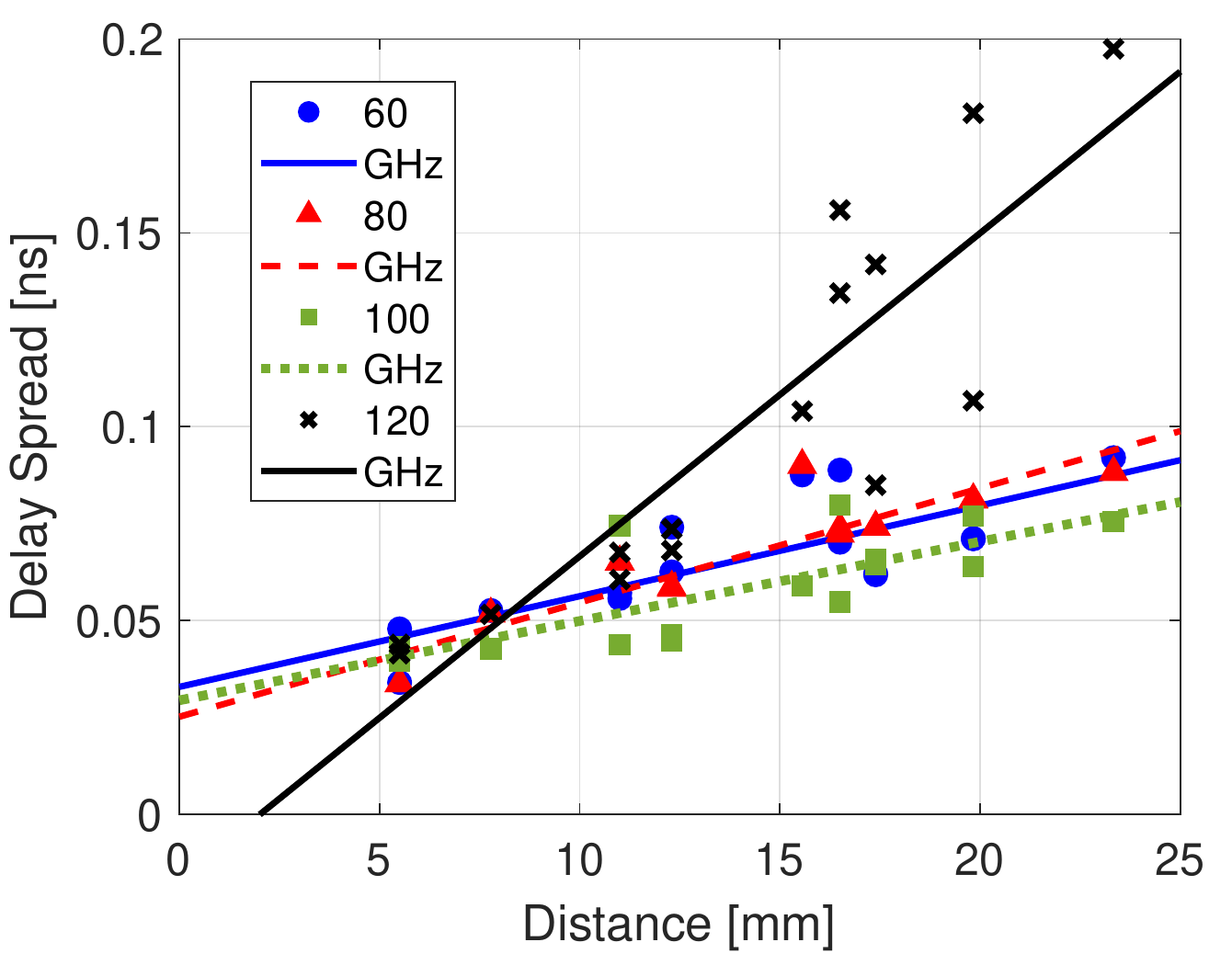}}
\subfigure[\label{fig:frequency}]{\includegraphics[width=0.32\textwidth]{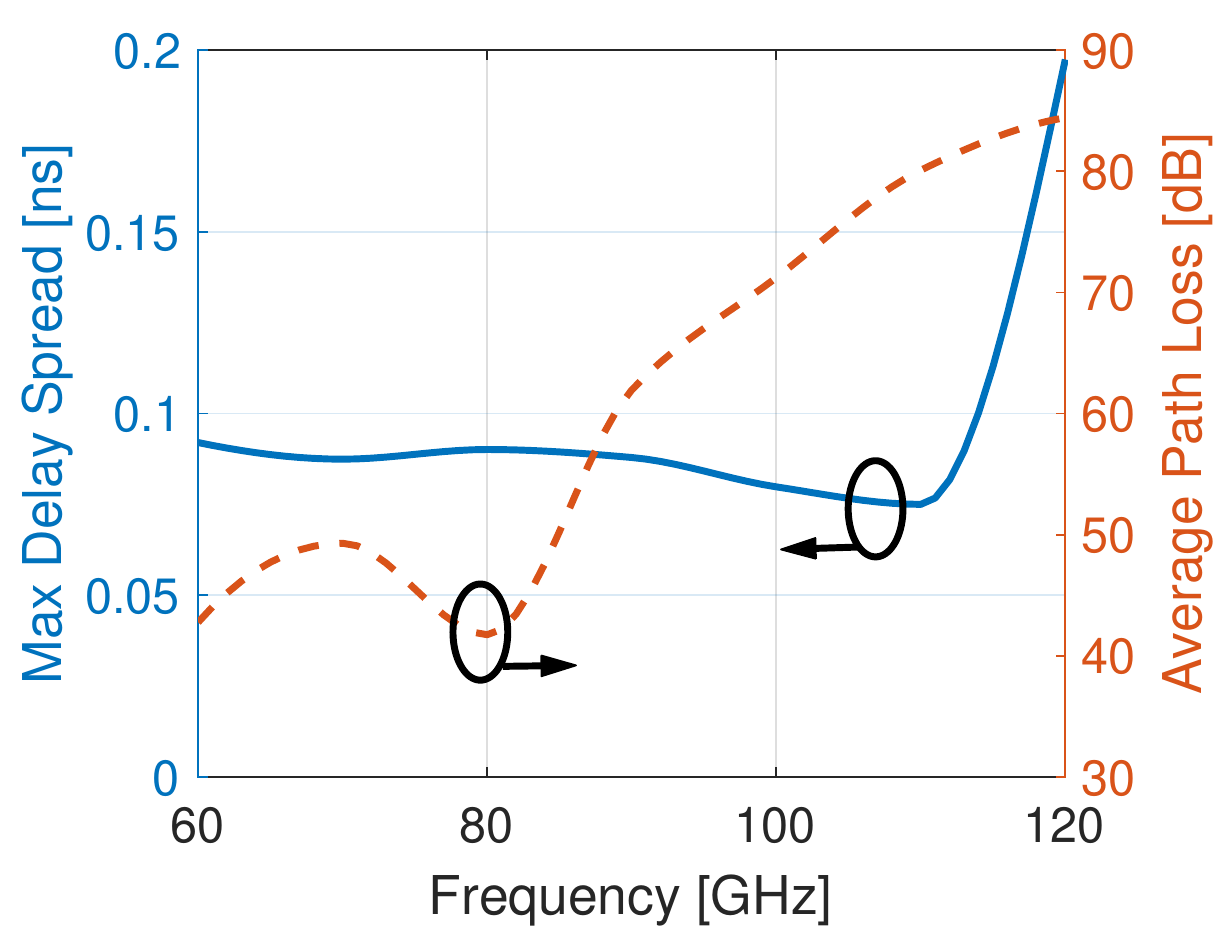}}
\vspace{-0.2cm}
\caption{Scaling analysis of the wireless channel with respect to the frequency $f$ for} $T_{s} = 0.3$ mm and $T_{h} = 0.8$ mm, detailing (a) path loss over distance, (b) delay spread over distance, and (c) maximum delay spread $\tau_{rms}$ and average path loss $L_{avg}$ as functions of $f$.\label{fig:frequencys}
\vspace{-0.3cm}
\end{figure*}


Finally, Figure \ref{fig:frequencys} presents the results of the frequency scaling analysis, which we limit to the 60--120 GHz span due to computational constraints. Additionally, we fix the silicon and heat spreader thicknesses to small and large values, respectively, following the design recommendations justified above. We chose this particular ($T_{s} = 0.3$ mm and $T_{h} = 0.8$ mm) because it is close to an optimal point with respect to dispersion. We find that $f_{c} = 110$ GHz leads to a minimum in terms of delay spread, although the improvement is limited with respect to the other frequencies. The impact on path loss, on the other hand, is substantial yet counter-intuitive at times as the average path loss drops first oscillates around 40--50 dB when shifting the frequency between 60 GHz and 90 GHz, to then increase substantially towards 90 dB at 120 GHz.

\subsection{Engineering the Channel} \label{sec:evalEng}
Here, we show the potential of channel engineering through a partial exploration of the $\{T_{h}, T_{s}, f_{c}\}$ design space. Our aim is not to fully implement the optimizer, but rather to validate the potential of the approach by confirming both the complex interactions between inputs and the presence of local optima, as well as by giving good approximations of the path loss and delay spread improvements that we can expect. 

\begin{figure}[!t]
\centering
\includegraphics[width=0.3\columnwidth]{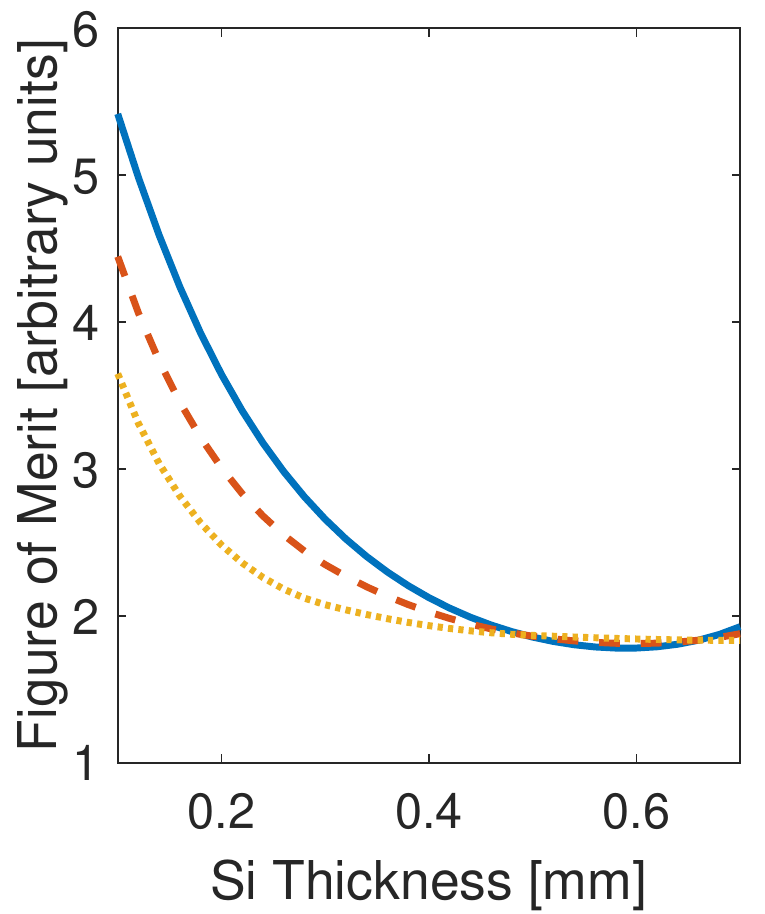}
\includegraphics[width=0.3\columnwidth]{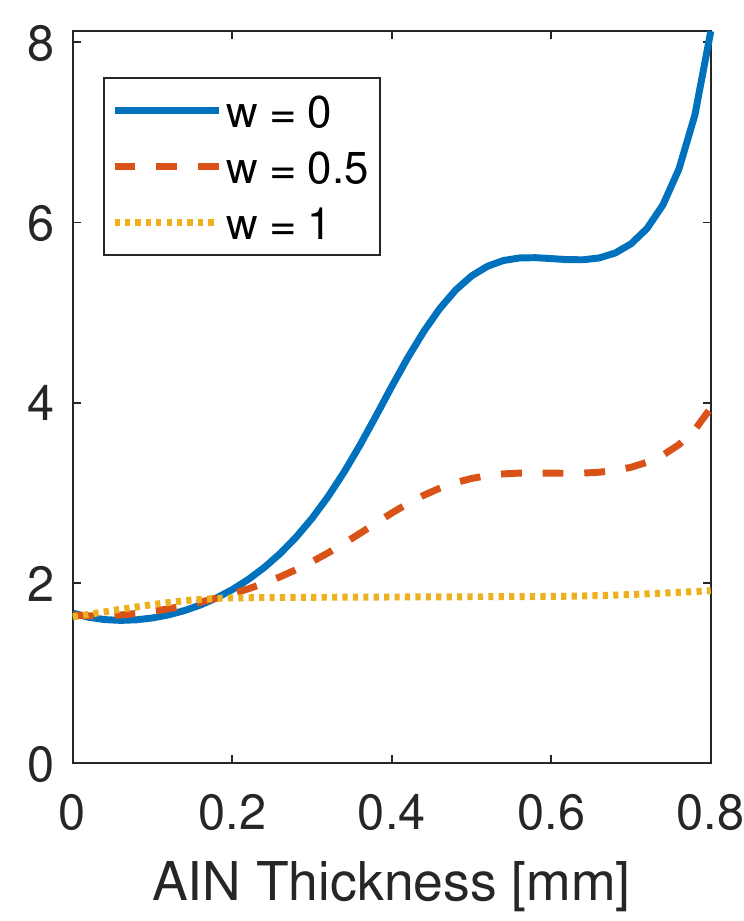}
\includegraphics[width=0.3\columnwidth]{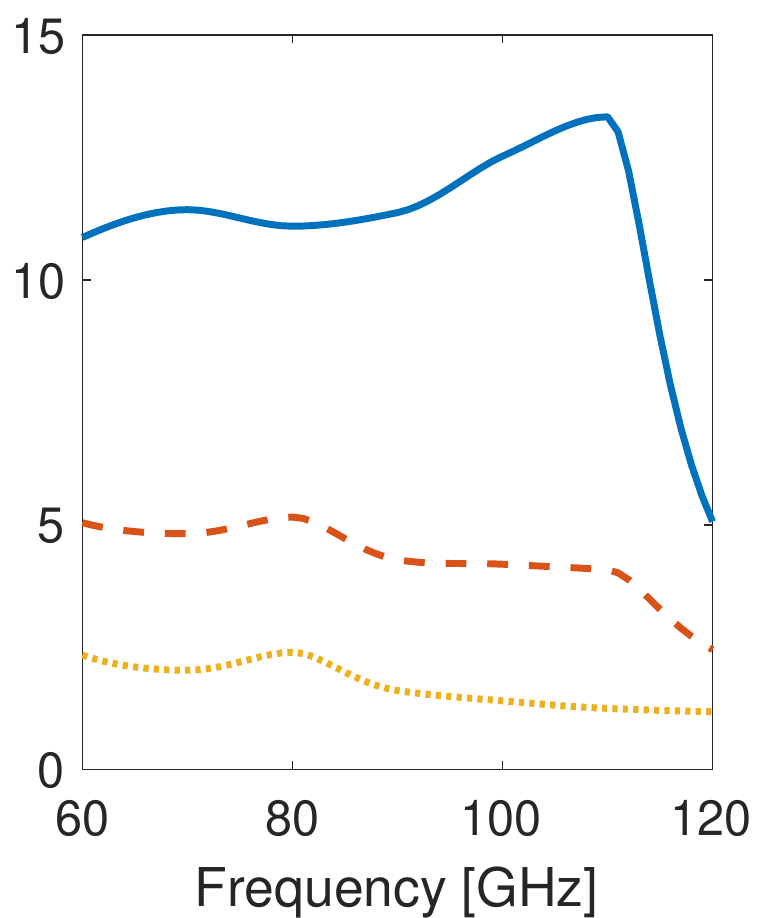}
\vspace{-0.2cm}
\caption{Figure of merit $\phi_{w}$ as function of $\{T_{s}, T_{h}, f_{c}\}$ for different priority weights. Unless noted, $T_{s} = 0.2$ mm, $T_{h} = 0.7$ mm, and $f_{c} = 60$ GHz.}
\vspace{-0.3cm}
\label{fig:optimizer}
\end{figure} 

We first plot the figure of merit $\phi_{w}$ as function of each exploration parameter while leaving the others fixed. The results, summarized in Figure \ref{fig:optimizer}, confirm the main lessons learned in Section \ref{sec:charact_eval}: thin silicon is generally preferable (left plot), it is hard to obtain clear tendencies with respect to the heat spreader (middle plot), and performance may plateau close to local optima (right plot). The choice of $w$ also plays an important role in the optimization and Figure \ref{fig:optimizer} also confirms it. Since path loss and delay spread often show opposed trends, the shape of $\phi$ changes in unexpected ways and causes wild variations in the optimal design points. Take, for instance, the frequency scaling trend. The optimal point is clearly at 110 GHz for $w=0$, but that peak dilutes progressively and disappears around $w=0.6$. At that point, the optimal frequency becomes 60 GHz or 80 GHz due to the better path loss behavior. 

\begin{figure}[!t]
\centering
\includegraphics[width=\columnwidth]{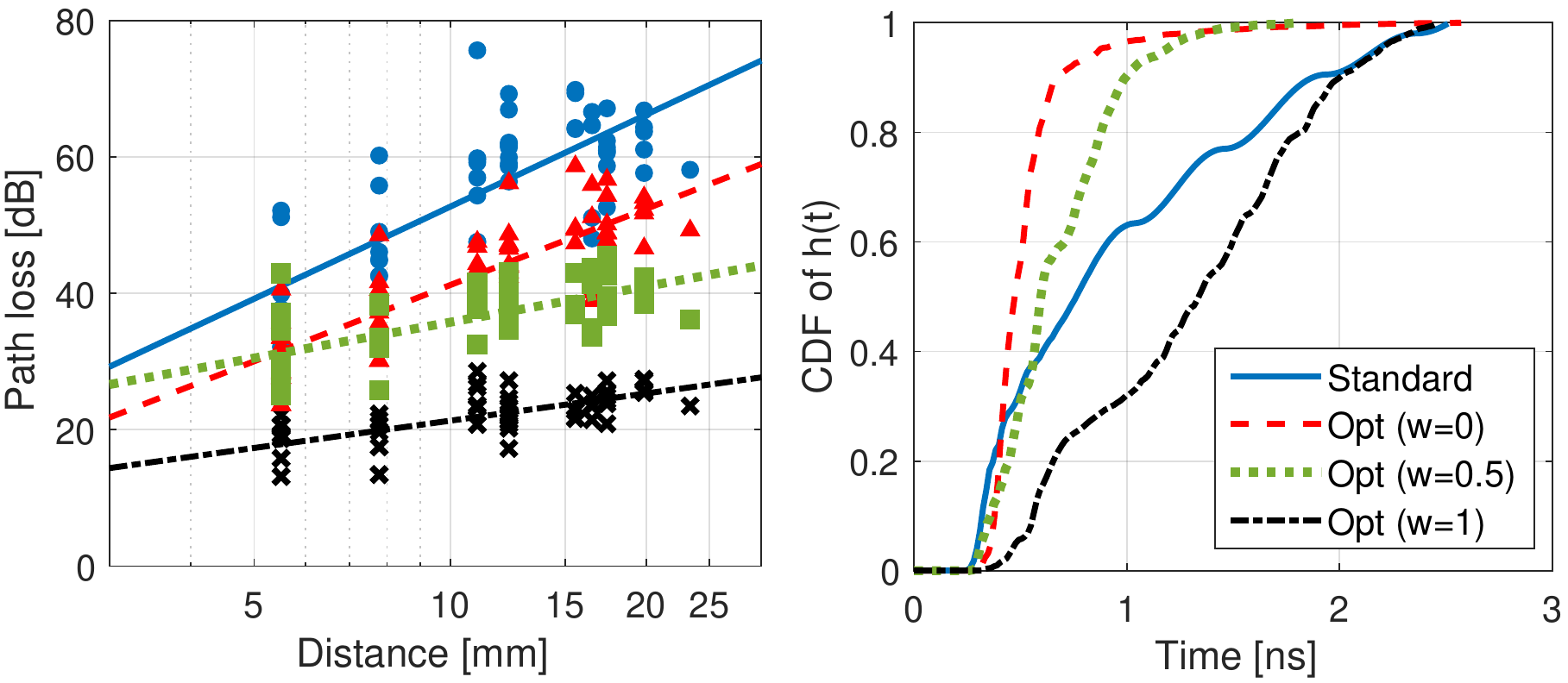}
\vspace{-0.6cm}
\caption{Comparison between standard package ($T_{s} = 0.7$ mm, $T_{h} = 0.2$ mm, $f_{c} = 60$ GHz) and optimal points for three different power--speed weights from the path loss (left) and delay spread perspectives (right).}
\vspace{-0.3cm}
\label{fig:optimized}
\end{figure}

In order to estimate the maximum gains that we can achieve through channel engineering, we further explored the design space in the quest for points close to a hypothetical global optima. We chose three representative values of $w$ and compared the results with those of a standard chip ($T_{s} = 0.7$ mm, $T_{h} = 0.2$ mm, $f_{c} = 60$ GHz). Figure \ref{fig:optimized} and Table \ref{tab:optimal} illustrate the outcome of this process. There $L_{max}$ and $L_{avg}$ refer to the maximum and average path loss across all measured transmitter-receiver pairs within the 4$\times$4 homogeneous grid of antennas.

We first set $w=0$ to simulate the extreme of high performance, thereby pushing the limits on the delay spread. The peak has been found around $\{T_{s} = 0.3$ mm, $T_{h} = 0.8$ mm, $f_{c} = 110$ GHz$\}$ and yields a worst-case delay spread of $\tau_{rms} = 71.32$ ps for a coherence bandwidth of $B_{c} = 14.02$ GHz. This is roughly one order of magnitude better than the standard chip case (0.52 ns for 1.92 GHz) and confirms that the speeds assumed in the \ac{WNoC} literature are feasible. In terms of path loss, this design point is also 10--15 dB better than the standard.

A second representative case would be $w=1$, which pushes the limits on the path loss. The peak has been found by thinning the silicon down to our lower limit and using a thick spreader: $\{T_{s} = 0.1$ mm, $T_{h} = 0.8$ mm, $f_{c} = 60$ GHz$\}$. This case achieves an outstanding path loss reduction of 47.07 dB for $L_{max}$ and 32.69 dB for $L_{avg}$ ($n=1.32$). Further, this confirms that the path loss figures assumed in the literature, around 25--35 dB, are indeed achievable even in the presence of a chip package. However, the delay spread is maintained at the levels of the standard chip in this case.

Finally, let $w=0.5$ to model a channel engineering process searching a balance between power and performance. In this case, a local peak has been found around the point $\{T_{s} = 0.1$ mm, $T_{h} = 0.38$ mm, $f_{c} = 70$ GHz$\}$. With respect to the standard chip, this design allows to improve the coherence bandwidth $B_{c}$ by 3.52$\times$ and the average path loss $L_{avg}$ by over 1.5$\times$. Although this may not be a global optimum, it illustrates the potential of the methodology.

\subsection{Static Transceiver Optimization} \label{sec:evalDesign}
Since we are interested in pushing the limits of performance, this section evaluates the transceiver improvements in the package engineered for high performance. Thus, we take the worst-case transient response of the $\{T_{s} = 0.3$ mm, $T_{h} = 0.8$ mm, $f_{c} = 110$ GHz$\}$ design point with a delay spread of $\tau_{rms} = 71.32$ ps. In all the studied cases, \ac{OOK}-modulated waveforms are convoluted with the transient response at the channel and fed to the receiver, which determines the hypothetical position of the next '0' or '1' symbol. The \ac{BER} is calculated assuming independent and equiprobable symbols.

\begin{figure*}[!t]
\centering
\subfigure[Scaling with SNR ($r_{b} = 10$ GHz, NRZ)\label{fig:threshold}]{\includegraphics[width=0.335\textwidth]{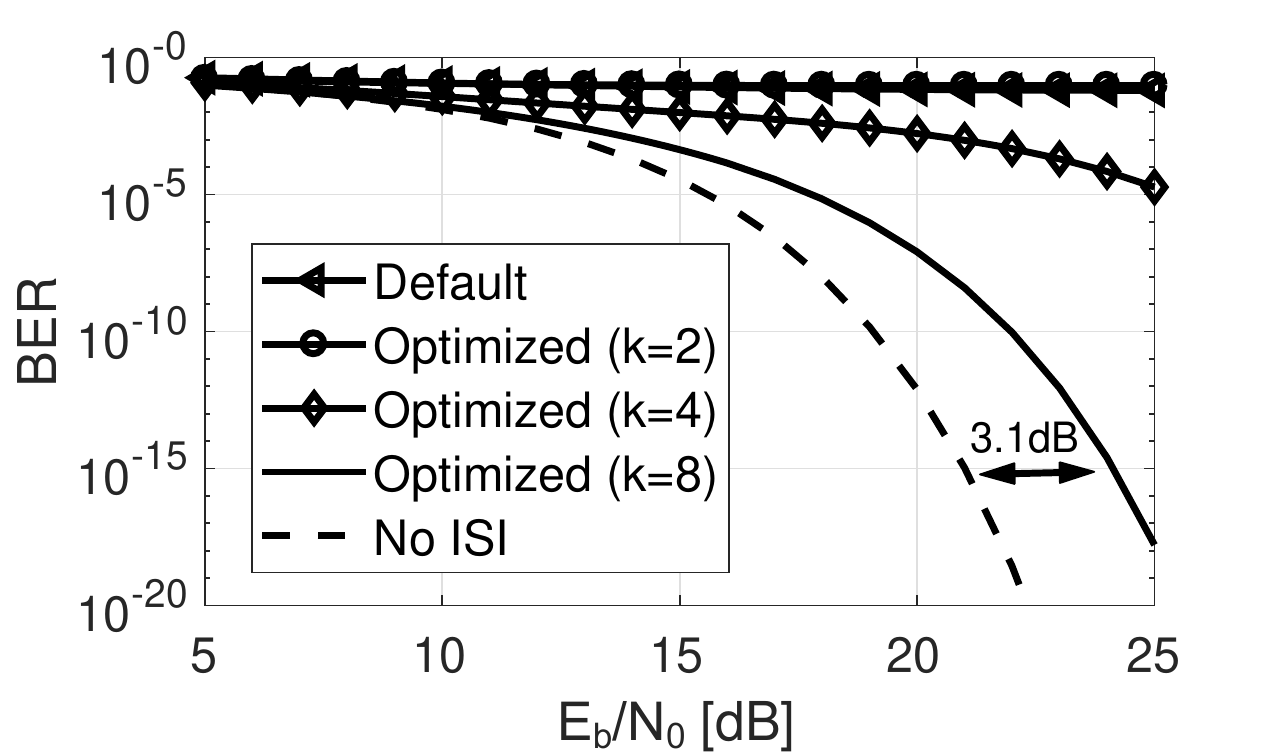}}
\subfigure[Scaling with bitrate ($P_{rx} = -52$ dBm, NRZ)\label{fig:threshold2}]{\includegraphics[width=0.31\textwidth]{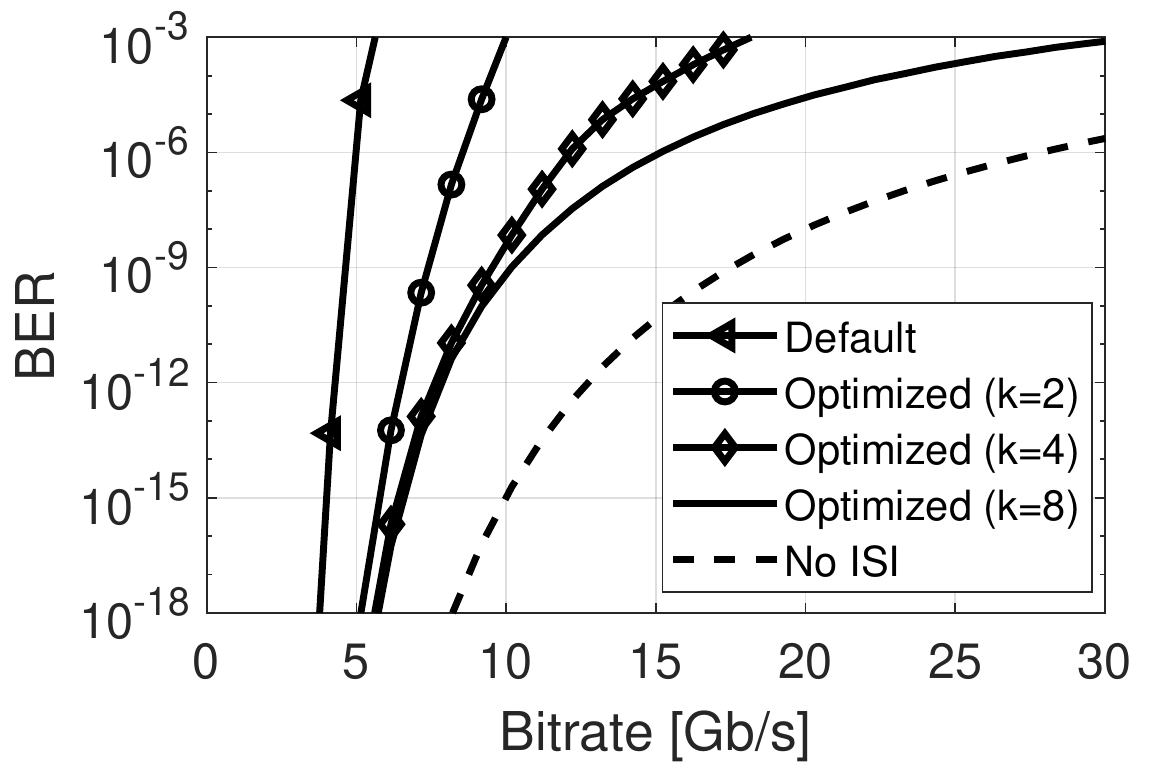}}
\subfigure[Scaling with RZ ($K=8$, $E_{b}/N_{0} = 21$ dB)\label{fig:RZ}]{\includegraphics[width=0.335\textwidth]{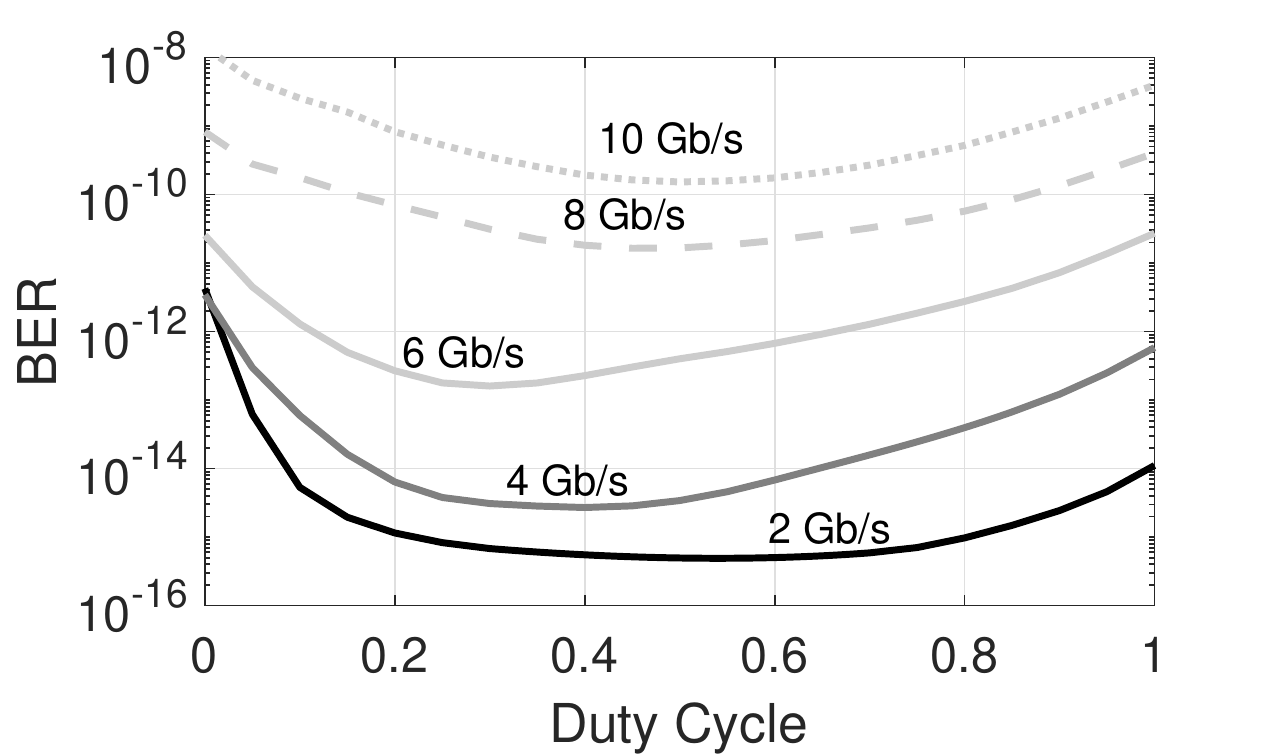}}
\vspace{-0.1cm}
\caption{Impact of transceiver optimizations on the Bit Error Rate (BER) assuming \ac{OOK} modulation. NRZ stands for Non-Return-to-Zero.}
\vspace{-0.2cm}
\label{fig:transceiverImpr}
\end{figure*}

\textbf{Threshold adaptation:} We simulate our proposed receiver with different number of decision thresholds $K$. We first obtain the threshold values by looking at the previous $\log_{2}(K)$ symbols and then use conventional \emph{erfc} formulation to derive the error probability. Figure \ref{fig:threshold} plots the resulting \ac{BER} for a fixed $r_{b}$ of 10 Gb/s, assumed in numerous \ac{WNoC} works, and as a function of $E_{b}/N_{0}$. Although we are below the coherence bandwidth, \ac{ISI} effects disable the use of \emph{a priori} thresholds based on steady state measurements alone. The performance for $K=4$ is far from ideal, but starts to improve significantly. At $K=8$, the receiver performs close to a coherent receiver in an \ac{ISI}-free environment. In fact, it only needs to be 24.1 dB above the noise floor achieve the stringent \ac{BER} required for \ac{WNoC} (10\textsuperscript{-15}). This is only 3.1 dB over the ideal case.

To further evaluate the 
potential of the proposed scheme, we fix the received power and push the data rate way beyond the coherence bandwidth. The results, shown in Figure \ref{fig:threshold2}, reveal that the receiver by default stops working upon reaching the \ac{ISI} wall at around 5 Gb/s. With as few as $K=2$ thresholds, our proposed scheme improves the achievable data rate between 20\% and 40\%. Again, increasing the number of decision thresholds allows to further mitigate \ac{ISI} (the bitrate increases from 7.32 up to 10.56 Gb/s at $BER = 10^{-9}$), to the point of becoming indispensable as we keep pushing the data rate. These results illustrate the tradeoff between performance and receiver complexity, although the overhead of our proposed scheme is arguably small.

\begin{table}[!t] 
\caption{Summary of the optimized package designs}
\vspace{-0.1cm}
\label{tab:optimal}
\footnotesize
\centering
\begin{tabular}{c|cc|ccc} 
 & $\tau_{rms}$ (ns) & $B_{c}$ (GHz) & $L_{max}$ (dB) & $L_{avg}$ (dB) & $n$ \\
\hline
$w=0$ & 0.07 & 14.02 & 58.62 & 42.76 & 3.28 \\ 
$w=0.5$ & 0.15 & 6.76 & 45.49 &	36.48 & 1.74 \\
$w=1$ & 0.59 & 1.69  & 28.55 & 21.88 & 1.32  \\ 
\hline
$Std.$ & 0.52 & 1.92 & 75.62 & 54.57 & 4.61 \\
\end{tabular}
\vspace{-0.5cm}
\end{table}

\textbf{Return-to-zero:} One of the conclusions that can be extracted from Figures \ref{fig:threshold} and \ref{fig:threshold2} is that we can minimize \ac{ISI}, but we cannot get rid of it completely. The adaptive threshold moves along with the average received energy, but cannot eliminate the case where the '0' and '1' symbols move closer. This is precisely the case targeted by \ac{RZ}. To evaluate it, we assume a receiver with $K=8$ and set the $E_{b}/N_{0}$ for all transmission speeds. The results, plotted in Figure \ref{fig:RZ}, demonstrate that there is indeed a duty cycle value that minimizes the error rate. The optimal point depends on the transmission speed and yields an improvement of up to two orders of magnitude with respect to non-RZ. The $E_{b}/N_{0}$ scaling analysis, not shown here in the interest of space, also revealed that \ac{RZ} brings our scheme 1.2 dB closer to the ideal receiver for $BER = 10^{-15}$.



\section{Discussion} 
\label{sec:discussion}
\textbf{Impact on transmission speed.} 
The channel engineering process, by means of substantial delay spread cuts, increases the \ac{ISI}-free speed by an order of magnitude with respect to in a standard chip. Further, the transceiver optimizations have demonstrated that (i) achieving a BER of 10\textsuperscript{-15} at 10 GHz is affordable, and that (ii) it would be otherwise impossible. This thereby proves that our methodology enables the speeds generally assumed in the \ac{WNoC} literature. 

\textbf{Impact on power consumption.} 
By reducing the path loss by up to 47 dB, we achieve attenuation levels close to those assumed in recent transceiver proposals (26.5 dB in \cite{Yu2014a, Yu2015} and 26 dB calculated with data in \cite{Subramaniam2017, Ahmed2018}). Meeting such assumptions would lead to a bit energies of 1.95 pJ/bit for \cite{Yu2014a, Yu2015} or 0.54 pJ/bit for \cite{Subramaniam2017, Ahmed2018}, along the lines of what is assumed in the \ac{WNoC} literature. On top of that, our transceiver only needs an extra 3.1 dB of SNR to compensate for the \ac{ISI} effects at 10 Gb/s and $BER = 10^{-15}$.

To make an explicit connection between channel losses and efficiency, we note that power amplifiers are the most consuming components of current transceivers, e.g., 70.8\% in \cite{Subramaniam2017, Ahmed2018}. Compensating for extra losses, noise figures, or circuit limitations would make these figures to increase even further. In fact, each amplifier has a limit $P_{sat}$ on the output power it can provide. Going beyond that limit would require a re-design of the amplifier and, according to long-time experimentally validated scaling tendencies, the extra effort is generally paid with a reduction of the amplifier efficiency in 2.5\% per each extra dBm of $P_{sat}$ \cite{PAs}. On this same direction, it is worth noting that increasing the frequency may impact not only on the channel, but also on the efficiency of the transceiver. Although there is no technological impediment preventing the use of the 60--120 GHz band (current technologies offer $f_{T}$ and $f_{max}$ values around 300 GHz and above), pushing the frequency may initially lead to a loss of efficiency. This difference, however, levels out as technology matures and its use is extended.



\textbf{Generality of results:} we note that the specific results contained in this paper are, by definition, valid for a particular chip arrangement and cannot be generalized to any chip package. The key takeaways of the present work are, however, that the wireless intra-chip channel can be optimized and that the proposed methodology is applicable to any chip package. Such optimization process is unique to this wireless communications scenario.

\textbf{Research directions:} although this work has mitigated the intra-chip channel impairments significantly, we do not consider to have reached a lower bound. Besides the application of simulated annealing techniques to find global optima, we could improve propagation further by (i) directing certain rays via reflectors or leveraging the multiple antennas already in place to perform beamforming, (ii) thinning silicon down to the manufacturing limits \cite{Bieck2010}, or (iii) exploring frequencies up to the terahertz band \cite{Chen2019}. Additionally, factors such as the chip's lateral dimensions, the antenna placement, or the resistivity of the silicon substrate \cite{Zhang2007} could be brought into the optimization process as long as the computational cost is affordable. At the transceiver side, low-weight coding would help minimizing the impact of \ac{ISI} at very high speeds \cite{Jornet2011OOK}. Further,  compact and efficient Forward Error Correction (FEC) techniques could allow to reach the required BER without placing a large burden on the amplifiers \cite{Chang2010, Ganguly2011}.

\textbf{Alternative technologies:} Transmission
of optical signals through integrated nanophotonic waveguides \cite{Batten2012, Sun2015, Kurian2010} or of RF signals through transmission lines (TLs) \cite{Carpenter2011, Oh2013} can provide low latency and broadcast. Compared to wireless intra-chip communication, both nanophotonics and TLs are more energy efficient and provide higher bandwidth, because energy is guided rather than radiated. In this respect, the present works aims to reduce the performance and efficiency gap with respect to its alternatives. Beyond that, the main downturn of nanophotonics and TLs is the need of a physical infrastructure to interconnect the nodes, which complicates the network design. Further, nanophotonics are less scalable due to laser power needs. Light is modulated by the transmitter and then guided to all the receivers. Each receiver extracts a fraction of the light, causing losses, and requiring high laser power for large destinations sets \cite{AbadalONDM}. On the other hand, TLs are less scalable due to the need for amplifiers along the transmission line and centralized arbitration, issues that are exacerbated if the fan-out is large. 

\section{Conclusion}
\label{sec:conclusion}
Wireless intra-chip communication has been proposed as a potential solution to the scalability problems of current multicore processors. However, we have demonstrated that most works on this field are overly optimistic with regards to the channel, assuming figures one or two orders of magnitude better than what we found for a standard chip package. To further address this fundamental issue and restate the potential of \ac{WNoC}, we proposed a methodology that exploits two unique traits of this new wireless scenario: its monolithic and static nature. The first allows us to engineer the channel, this is, to modify the chip package to enhance propagation in manufacturer-friendly ways. This process is applicable to any chip package and, here, we have illustrated its potential by showing improvements of 47 dB of path loss or more than 10 GHz in coherence bandwidth for a particular system. The second allows us to optimize the transceiver to mitigate multipath effects beyond the Nyquist limit. We demonstrated that we can decode \ac{OOK} signals at 10 Gb/s with a BER of 10\textsuperscript{-15} with a signal-to-noise ratio only 3.1 dB greater than in a dispersion-free environment.





\bibliographystyle{IEEEtran}
\bibliography{IEEEabrv,references,new-refs}

\begin{thebibliography}{10}
\providecommand{\url}[1]{#1}
\csname url@samestyle\endcsname
\providecommand{\newblock}{\relax}
\providecommand{\bibinfo}[2]{#2}
\providecommand{\BIBentrySTDinterwordspacing}{\spaceskip=0pt\relax}
\providecommand{\BIBentryALTinterwordstretchfactor}{4}
\providecommand{\BIBentryALTinterwordspacing}{\spaceskip=\fontdimen2\font plus
\BIBentryALTinterwordstretchfactor\fontdimen3\font minus
  \fontdimen4\font\relax}
\providecommand{\BIBforeignlanguage}[2]{{%
\expandafter\ifx\csname l@#1\endcsname\relax
\typeout{** WARNING: IEEEtran.bst: No hyphenation pattern has been}%
\typeout{** loaded for the language `#1'. Using the pattern for}%
\typeout{** the default language instead.}%
\else
\language=\csname l@#1\endcsname
\fi
#2}}
\providecommand{\BIBdecl}{\relax}
\BIBdecl

\bibitem{Marculescu2009}
R.~Marculescu, U.~Ogras, L.-S. Peh, N.~{Enright Jerger}, and Y.~Hoskote,
  ``{Outstanding research problems in NoC design: system, microarchitecture,
  and circuit perspectives},'' \emph{IEEE Transactions on Computer-Aided Design
  of Integrated Circuits and Systems}, vol.~28, no.~1, pp. 3--21, 2009.

\bibitem{Vangal2008}
S.~Vangal, J.~Howard, G.~Ruhl, S.~Dighe, H.~Wilson, J.~Tschanz, D.~Finan,
  A.~Singh, T.~Jacob, S.~Jain, V.~Erraguntla, C.~Roberts, Y.~Hoskote,
  N.~Borkar, and S.~Borkar, ``{An 80-Tile Sub-100-W TeraFLOPS Processor in
  65-nm CMOS},'' \emph{IEEE Journal of Solid-State Circuits}, vol.~43, no.~1,
  pp. 29--41, 2008.

\bibitem{Nychis2012}
G.~Nychis, C.~Fallin, and T.~Moscibroda, ``{On-chip networks from a networking
  perspective: congestion and scalability in many-core interconnects},'' in
  \emph{Proceedings of the SIGCOMM}, 2012, pp. 407--18.

\bibitem{Park2012a}
S.~Park, T.~Krishna, C.-H. Chen, B.~Daya, A.~Chandrakasan, and L.-S. Peh,
  ``{Approaching the theoretical limits of a mesh NoC with a 16-node chip
  prototype in 45nm SOI},'' in \emph{Proceedings of the DAC-49}, 2012, pp.
  398--405.

\bibitem{Chen2015a}
G.~Chen, M.~A. Anders, H.~Kaul, S.~K. Satpathy, S.~K. Mathew, S.~K. Hsu,
  A.~Agarwal, R.~K. Krishnamurthy, V.~De, and S.~Borkar, ``{A 340 mV-to-0.9 v
  20.2 Tb/s source-synchronous hybrid packet/circuit-switched 16 x 16
  network-on-chip in 22 nm tri-gate CMOS},'' \emph{IEEE Journal of Solid-State
  Circuits}, vol.~50, no.~1, pp. 59--67, 2015.

\bibitem{Wentzlaff2007}
D.~Wentzlaff, P.~Griffin, H.~Hoffmann, L.~Bao, B.~Edwards, C.~Ramey,
  M.~Mattina, C.-C. Miao, J.~F. {Brown III}, and A.~Agarwal, ``{On-chip
  interconnection architecture of the tile processor},'' \emph{IEEE Micro},
  vol.~27, no.~5, pp. 15--31, 2007.

\bibitem{XeonPhi}
G.~Chrysos, ``Intel{\textregistered} xeon phi™ coprocessor-the
  architecture,'' \emph{Intel Whitepaper}, vol. 176, 2014.

\bibitem{Bertozzi2014}
D.~Bertozzi, G.~Dimitrakopoulos, J.~Flich, and S.~Sonntag, ``{The fast evolving
  landscape of on-chip communication},'' \emph{Design Automation for Embedded
  Systems}, vol.~19, no.~1, pp. 59--76, 2015.

\bibitem{Kim2012Survey}
J.~Kim, K.~Choi, and G.~Loh, ``{Exploiting new interconnect technologies in
  on-chip communication},'' \emph{IEEE Journal on Emerging and Selected Topics
  in Circuits and Systems}, vol.~2, no.~2, pp. 124--136, 2012.

\bibitem{Matolak2012}
D.~Matolak, A.~Kodi, S.~Kaya, D.~DiTomaso, S.~Laha, and W.~Rayess, ``{Wireless
  networks-on-chips: architecture, wireless channel, and devices},'' \emph{IEEE
  Wireless Communications}, vol.~19, no.~5, 2012.

\bibitem{AbadalMICRO}
S.~Abadal, B.~Sheinman, O.~Katz, O.~Markish, D.~Elad, Y.~Fournier, D.~Roca,
  M.~Hanzich, G.~Houzeaux, M.~Nemirovsky, E.~Alarc{\'{o}}n, and
  A.~Cabellos-Aparicio, ``{Broadcast-Enabled Massive Multicore Architectures: A
  Wireless RF Approach},'' \emph{IEEE Micro}, vol.~35, no.~5, pp. 52--61, 2015.

\bibitem{Kim2016}
R.~G. Kim, W.~Choi, Z.~Chen, P.~P. Pande, D.~Marculescu, and R.~Marculescu,
  ``{Wireless NoC and Dynamic VFI Codesign: Energy Efficiency Without
  Performance Penalty},'' \emph{IEEE Transactions on Very Large Scale
  Integration (VLSI) Systems}, vol.~24, no.~7, pp. 2488--2501, 2016.

\bibitem{Sikder2016}
M.~A.~I. Sikder, A.~Kodi, W.~Rayess, D.~Ditomaso, D.~Matolak, and S.~Kaya,
  ``{Exploring wireless technology for off-chip memory access},'' in
  \emph{Proceedings of the HOTI '16}, 2016, pp. 92--99.

\bibitem{Markish2015}
O.~Markish, B.~Sheinman, O.~Katz, D.~Corcos, and D.~Elad, ``{On-chip mmWave
  Antennas and Transceivers},'' in \emph{Proceedings of the NoCS '15}, 2015, p.
  Art. 11.

\bibitem{Cheema2013}
H.~M. Cheema and A.~Shamim, ``{The last barrier: On-chip antennas},''
  \emph{IEEE Microwave Magazine}, vol.~14, no.~1, pp. 79--91, 2013.

\bibitem{Wu2017b}
J.~Wu, A.~Kodi, S.~Kaya, A.~Louri, and H.~Xin, ``{Monopoles Loaded with
  3-D-Printed Dielectrics for Future Wireless Intra-Chip Communications},''
  \emph{IEEE Transactions on Antennas and Propagation}, vol.~65, no.~12, pp.
  6838--6846, 2017.

\bibitem{Floyd2002}
B.~A. Floyd, C.-M. Hung, and K.~K. O, ``{Intra-chip wireless interconnect for
  clock distribution implemented with integrated antennas, receivers, and
  transmitters},'' \emph{IEEE Journal of Solid-State Circuits}, vol.~37, no.~5,
  pp. 543--552, 2002.

\bibitem{Zhang2007}
Y.~P. Zhang, Z.~M. Chen, and M.~Sun, ``{Propagation Mechanisms of Radio Waves
  Over Intra-Chip Channels With Integrated Antennas: Frequency-Domain
  Measurements and Time-Domain Analysis},'' \emph{IEEE Transactions on Antennas
  and Propagation}, vol.~55, no.~10, pp. 2900--2906, 2007.

\bibitem{Wu2013a}
H.-T. Wu, J.-J. Lin, and K.~K. O, ``{Inter-Chip Wireless Communication},'' in
  \emph{Proceedings of the EuCAP '13}, 2013, pp. 3647--3649.

\bibitem{Yu2014}
X.~Yu, J.~Baylon, P.~Wettin, D.~Heo, P.~{Pratim Pande}, and S.~Mirabbasi,
  ``{Architecture and Design of Multi-Channel Millimeter-Wave Wireless
  Network-on-Chip},'' \emph{IEEE Design {\&} Test}, vol.~31, no.~6, pp. 19--28,
  2014.

\bibitem{Laha2015}
S.~Laha, S.~Kaya, D.~W. Matolak, W.~Rayess, D.~DiTomaso, and A.~Kodi, ``{A New
  Frontier in Ultralow Power Wireless Links: Network-on-Chip and Chip-to-Chip
  Interconnects},'' \emph{IEEE Transactions on Computer-Aided Design of
  Integrated Circuits and Systems}, vol.~34, no.~2, pp. 186--198, 2015.

\bibitem{AbadalTON}
S.~Abadal, M.~Iannazzo, M.~Nemirovsky, A.~Cabellos-Aparicio, H.~Lee, and
  E.~Alarc{\'{o}}n, ``{On the Area and Energy Scalability of Wireless
  Network-on-Chip: A Model-based Benchmarked Design Space Exploration},''
  \emph{IEEE/ACM Transactions on Networking}, vol.~23, no.~5, pp. 1501--13,
  2015.

\bibitem{Mineo2015}
A.~Mineo, M.~Palesi, G.~Ascia, and V.~Catania, ``{Runtime Tunable Transmitting
  Power Technique in mm-Wave WiNoC Architectures},'' \emph{IEEE Transactions on
  VLSI Systems}, vol.~24, no.~4, pp. 1535--1545, 2016.

\bibitem{Abadal2018}
S.~Abadal, J.~Torrellas, E.~Alarc{\'{o}}n, and A.~Cabellos-Aparicio,
  ``{OrthoNoC: A Broadcast-Oriented Dual-Plane Wireless Network-on-Chip
  Architecture},'' \emph{IEEE Transactions on Parallel and Distributed
  Systems}, vol.~29, no.~3, pp. 628--641, 2018.

\bibitem{Sujay2012}
S.~Deb, A.~Ganguly, P.~P. Pande, B.~Belzer, and D.~Heo, ``{Wireless NoC as
  Interconnection Backbone for Multicore Chips: Promises and Challenges},''
  \emph{IEEE Journal on Emerging and Selected Topics in Circuits and Systems},
  vol.~2, no.~2, pp. 228--239, 2012.

\bibitem{Gade2017a}
S.~H. Gade and S.~Deb, ``{HyWin: Hybrid wireless NoC with sandboxed
  sub-networks for CPU/GPU architectures},'' \emph{IEEE Transactions on
  Computers}, vol.~66, no.~7, pp. 1145--1158, 2017.

\bibitem{DiTomaso2015}
D.~DiTomaso, A.~Kodi, D.~Matolak, S.~Kaya, S.~Laha, and W.~Rayess, ``{A-WiNoC:
  Adaptive Wireless Network-on-Chip Architecture for Chip Multiprocessors},''
  \emph{IEEE Transactions on Parallel and Distributed Systems}, vol.~26,
  no.~12, pp. 3289--3302, 2015.

\bibitem{Choi2018}
W.~Choi, K.~Duraisamy, R.~G. Kim, J.~R. Doppa, P.~P. Pande, D.~Marculescu, and
  R.~Marculescu, ``{On-Chip Communication Network for Efficient Training of
  Deep Convolutional Networks on Heterogeneous Manycore Systems},'' \emph{IEEE
  Transactions on Computers}, vol.~67, no.~5, pp. 672--686, 2018.

\bibitem{Abadal2018a}
S.~Abadal, A.~Mestres, J.~Torrellas, E.~Alarc{\'{o}}n, and
  A.~Cabellos-Aparicio, ``{Medium Access Control in Wireless Network-on-Chip: A
  Context Analysis},'' \emph{IEEE Communications Magazine}, vol.~56, no.~6, pp.
  172--178, 2018.

\bibitem{abadal2019wave}
S.~Abadal, C.~Han, and J.~M. Jornet, ``Wave propagation and channel modeling in
  chip-scale wireless communications: A survey from millimeter-wave to
  terahertz and optics,'' \emph{IEEE Access}, vol.~8, pp. 278--293, 2019.

\bibitem{Yan2009}
L.~Yan and G.~W. Hanson, ``{Wave propagation mechanisms for intra-chip
  communications},'' \emph{IEEE Transactions on Antennas and Propagation},
  vol.~57, no.~9, pp. 2715--2724, 2009.

\bibitem{Chen2009}
W.-H. Chen, S.~Joo, S.~Sayilir, R.~Willmot, T.-Y. Choi, D.~Kim, J.~Lu,
  D.~Peroulis, and B.~Jung, ``{A 6-Gb/s Wireless Inter-Chip Data Link Using
  43-GHz Transceivers and Bond-Wire Antennas},'' \emph{IEEE Journal of
  Solid-State Circuits}, vol.~44, no.~10, pp. 2711--2721, oct 2009.

\bibitem{Yeh2013}
H.~H. Yeh, N.~Hiramatsu, and K.~L. Melde, ``{The design of broadband 60 GHz AMC
  antenna in multi-chip RF data transmission},'' \emph{IEEE Transactions on
  Antennas and Propagation}, vol.~61, no.~4, pp. 1623--1630, 2013.

\bibitem{Narde2019}
R.~S. Narde, J.~Venkataraman, A.~Ganguly, and I.~Puchades, ``{Intra-and
  Inter-Chip Transmission of Millimeter-Wave Interconnects in NoC-based
  Multi-Chip Systems},'' \emph{IEEE Access}, vol.~7, pp. 112\,200--15, 2019.

\bibitem{Gade2017}
S.~H. Gade, S.~Garg, and S.~Deb, ``{OFDM Based High Data Rate, Fading Resilient
  Transceiver for Wireless Networks-on-Chip},'' in \emph{Proceedings of the
  ISVLSI '17}, 2017, pp. 483--488.

\bibitem{Rayess2017}
W.~Rayess, D.~W. Matolak, S.~Kaya, and A.~K. Kodi, ``{Antennas and Channel
  Characteristics for Wireless Networks on Chips},'' \emph{Wireless Personal
  Communications}, vol.~95, no.~4, pp. 5039--5056, 2017.

\bibitem{Elmasri2019}
I.~{El Masri}, T.~{Le Gouguec}, P.-M. Martin, R.~Allanic, and C.~Quendo,
  ``{Electromagnetic Characterization of the Intra-chip Propagation Channel in
  Ka and V Bands},'' \emph{IEEE Transactions on Components, Packaging and
  Manufacturing Technology}, vol.~9, no.~10, pp. 1931--1941, 2019.

\bibitem{AbadalASPLOS}
S.~Abadal, E.~Alarc{\'{o}}n, A.~Cabellos-Aparicio, and J.~Torrellas, ``{WiSync:
  An Architecture for Fast Synchronization through On-Chip Wireless
  Communication},'' in \emph{Proceedings of the ASPLOS '16}, 2016, pp. 3--17.

\bibitem{Fernando2019}
V.~Fernando, A.~Franques, S.~Abadal, S.~Misailovic, and J.~Torrellas,
  ``{Replica: A Wireless Manycore for Communication-Intensive and Approximate
  Data},'' in \emph{Proceedings of the ASPLOS '19}, 2019.

\bibitem{Yu2014a}
X.~Yu, S.~P. Sah, H.~Rashtian, S.~Mirabbasi, P.~P. Pande, and D.~Heo, ``{A
  1.2-pJ/bit 16-Gb/s 60-GHz OOK Transmitter in 65-nm CMOS for Wireless
  Network-On-Chip},'' \emph{IEEE Transactions on Microwave Theory and
  Techniques}, vol.~62, no.~10, pp. 2357--2369, 2014.

\bibitem{Yu2015}
X.~Yu, H.~Rashtian, and S.~Mirabbasi, ``{An 18.7-Gb/s 60-GHz OOK Demodulator in
  65-nm CMOS for Wireless Network-on-Chip},'' \emph{IEEE Transactions on
  Circuits And Systems -I: Regular Papers}, vol.~62, no.~3, pp. 799--806, 2015.

\bibitem{Subramaniam2017}
S.~Subramaniam, T.~Shinde, P.~Deshmukh, S.~Shamim, M.~Indovina, and A.~Ganguly,
  ``{A 0.36pJ/bit, 17Gbps OOK Receiver in 45-nm CMOS for Inter and Intra-Chip
  Wireless Interconnects},'' in \emph{Proceedings of the SOCC '17}, 2017.

\bibitem{Ahmed2018}
T.~Shinde, S.~Subramaniam, P.~Deshmukh, M.~M. Ahmed, M.~Indovina, and
  A.~Ganguly, ``{A 0.24 pJ/bit, 16 Gbps OOK Transmitter Circuit in 45-nm CMOS
  for Inter and Intra-Chip Wireless Interconnects},'' in \emph{Proceedings of
  the GLSVLSI '18}, 2018, pp. 69--74.

\bibitem{Matolak2013CHANNEL}
D.~Matolak, S.~Kaya, and A.~Kodi, ``{Channel modeling for wireless
  networks-on-chips},'' \emph{IEEE Communications Magazine}, vol.~51, no.~6,
  pp. 180--186, 2013.

\bibitem{ITRS}
\BIBentryALTinterwordspacing
``{ITRS: International Technology Roadmap for Semiconductors}.'' [Online].
  Available: \url{http://www.itrs2.net}
\BIBentrySTDinterwordspacing

\bibitem{Branch2005}
J.~Branch, X.~Guo, L.~Gao, A.~Sugavanam, J.~J. Lin, and K.~K. O, ``{Wireless
  communication in a flip-chip package using integrated antennas on silicon
  substrates},'' \emph{IEEE Electron Device Letters}, vol.~26, no.~2, pp.
  115--117, 2005.

\bibitem{Pano2019}
V.~Pano, I.~Tekin, Y.~Liu, K.~R. Dandekar, and B.~Taskin, ``{In-Package
  Wireless Communication with TSV-based Antenna},'' in \emph{Proceedings of the
  ISCAS '19}, 2019, pp. 19--21.

\bibitem{Timoneda2018}
X.~Timoneda, S.~Abadal, A.~Cabellos-Aparicio, D.~Manessis, J.~Zhou,
  A.~Franques, J.~Torrellas, and E.~Alarc{\'{o}}n, ``{Millimeter-Wave
  Propagation within a Computer Chip Package},'' in \emph{Proceedings of the
  ISCAS '18}, 2018.

\bibitem{Mondal2016}
H.~Mondal, S.~Gade, M.~Shamim, S.~Deb, and A.~Ganguly, ``{Interference-Aware
  Wireless Network-on-Chip Architecture using Directional Antennas},''
  \emph{IEEE Transactions on Multi-Scale Computing Systems}, vol.~3, no.~3, pp.
  193--205, 2017.

\bibitem{Wang2016AWPL}
W.~Wang, Y.~Chen, S.~Yang, Q.~Cao, H.~Li, X.~Zheng, and Y.~Wang, ``{Wireless
  inter/intra-chip communication using an innovative PCB channel bounded by a
  metamaterial absorber},'' \emph{IEEE Antennas and Wireless Propagation
  Letters}, vol.~15, pp. 1634--1637, 2016.

\bibitem{Kim2001}
K.~Kim, W.~Bornstad, and K.~K. O, ``{A Plane Wave Model Approach to
  Understanding Propagation in an Intra-chip Communication System},'' in
  \emph{Proceedings of the APS '01}, 2001, pp. 166--169.

\bibitem{Khademi2015}
S.~Khademi, S.~{Prabhakar Chepuri}, Z.~Irahhauten, G.~Janssen, and A.-J.
  van~der Veen, ``{Channel Measurements and Modeling for a 60 GHz Wireless Link
  Within a Metal Cabinet},'' \emph{IEEE Transactions on Wireless
  Communications}, vol.~14, no.~9, pp. 5098--5110, 2015.

\bibitem{Chiang2010}
P.~Y. Chiang, S.~Woracheewan, C.~Hu, L.~Guo, H.~Liu, R.~Khanna, and J.~Nejedlo,
  ``{Short-Range, Wireless Interconnect within a Computing Chassis: Design
  Challenges},'' \emph{IEEE Design {\&} Test of Computers}, vol.~27, no.~4, pp.
  32--43, 2010.

\bibitem{Zajic2018}
A.~Zajic and P.~Juyal, ``{Modeling of THz Chip-to-Chip Wireless Channels in
  Metal Enclosures},'' in \emph{Proceedings of the EuCAP '18}, 2018.

\bibitem{CST}
\BIBentryALTinterwordspacing
``{CST Microwave Studio}.'' [Online]. Available: \url{http://www.cst.com}
\BIBentrySTDinterwordspacing

\bibitem{Timoneda2018b}
X.~Timoneda, A.~Cabellos-Aparicio, D.~Manessis, E.~Alarc{\'{o}}n, and
  S.~Abadal, ``{Channel Characterization for Chip-scale Wireless Communications
  within Computing Packages},'' in \emph{Proceedings of the NOCS '18}, 2018.

\bibitem{Lin2007}
J.~Lin, H.~Wu, Y.~Su, L.~Gao, A.~Sugavanam, and J.~Brewer, ``{Communication
  using antennas fabricated in silicon integrated circuits},'' \emph{IEEE
  Journal of Solid-State Circuits}, vol.~42, no.~8, pp. 1678--1687, 2007.

\bibitem{nam2000multiobjective}
D.~Nam and C.~H. Park, ``Multiobjective simulated annealing: A comparative
  study to evolutionary algorithms,'' \emph{International Journal of Fuzzy
  Systems}, vol.~2, no.~2, pp. 87--97, 2000.

\bibitem{Bieck2010}
F.~Bieck, S.~Spiller, F.~Molina, M.~T{\"{o}}pper, C.~Lopper, I.~Kuna, T.~C.
  Seng, and T.~Tabuchi, ``{Carrierless design for handling and processing of
  ultrathin wafers},'' \emph{Proceedings of the ECTC '10}, pp. 316--322, 2010.

\bibitem{Simkin1992}
J.~Simkin and C.~W. Trowbridge, ``{Optimizing electromagnetic devices combining
  direct search methods with simulated annealing},'' \emph{IEEE Transactions on
  Magnetics}, vol.~28, no.~2, pp. 1545--1548, 1992.

\bibitem{Shu2004}
L.-S. S. L.-S. Shu, S.-Y. H. S.-Y. Ho, and S.~J. Ho, ``{A novel orthogonal
  simulated annealing algorithm for optimization of electromagnetic
  problems},'' pp. 1791--1795, 2004.

\bibitem{PAs}
\BIBentryALTinterwordspacing
H.~Wang, F.~Wang, H.~T. Nguyen, and S.~Li, ``{Power Amplifiers Performance
  Survey 2000-present}.'' [Online]. Available:
  \url{https://gems.ece.gatech.edu/PA{\_}survey.html}
\BIBentrySTDinterwordspacing

\bibitem{Chen2019}
Y.~Chen, X.~Cai, and C.~Han, ``{Wave Propagation Modeling for mmWave and
  Terahertz Wireless Networks-on-Chip Communications},'' in \emph{Proceedings
  of the ICC '19}.\hskip 1em plus 0.5em minus 0.4em\relax IEEE, 2019.

\bibitem{Jornet2011OOK}
J.~M. Jornet and I.~F. Akyildiz, ``{Low-Weight Channel Coding for Interference
  Mitigation in Electromagnetic Nanonetworks in the Terahertz Band},'' in
  \emph{Proceedings of the ICC '11}, 2011, pp. 1--6.

\bibitem{Chang2010}
F.~Chang, K.~Onohara, and T.~Mizuochi, ``{Forward error correction for 100 G
  transport networks},'' \emph{IEEE Communications Magazine}, vol.~48, no.~3,
  pp. 48--55, 2010.

\bibitem{Ganguly2011}
A.~Ganguly, P.~{Pratim Pande}, B.~Belzer, and A.~Nojeh, ``{A Unified Error
  Control Coding Scheme to Enhance the Reliability of a Hybrid Wireless
  Network-on-Chip},'' in \emph{Proceedings of the DFT '11}, 2011, pp. 277--285.

\bibitem{Batten2012}
C.~Batten, A.~Joshi, V.~Stojanovic, and K.~Asanovic, ``{Designing Chip-Level
  Nanophotonic Interconnection Networks},'' \emph{IEEE Journal on Emerging and
  Selected Topics in Circuits and Systems}, vol.~2, no.~2, pp. 137--153, 2012.

\bibitem{Sun2015}
C.~Sun, M.~T. Wade, Y.~Lee, J.~S. Orcutt, L.~Alloatti, M.~S. Georgas, A.~S.
  Waterman, J.~M. Shainline, R.~R. Avizienis, S.~Lin, B.~R. Moss, R.~Kumar,
  F.~Pavanello, A.~H. Atabaki, H.~M. Cook, A.~J. Ou, J.~C. Leu, Y.-H. Chen,
  K.~Asanovi{\'{c}}, R.~J. Ram, M.~A. Popovi{\'{c}}, and V.~M.
  Stojanovi{\'{c}}, ``{Single-chip microprocessor that communicates directly
  using light},'' \emph{Nature}, vol. 528, no. 7583, pp. 534--538, 2015.

\bibitem{Kurian2010}
G.~Kurian, J.~Miller, J.~Psota, J.~Eastep, J.~Liu, J.~Michel, L.~Kimerling, and
  A.~Agarwal, ``{ATAC: A 1000-Core Cache-Coherent Processor with On-Chip
  Optical Network},'' in \emph{Proceedings of the PACT}, 2010, pp. 477--488.

\bibitem{Carpenter2011}
A.~Carpenter, J.~Hu, J.~Xu, M.~Huang, and H.~Wu, ``{A case for globally
  shared-medium on-chip interconnect},'' in \emph{Proceedings of the ISCA-38},
  2011, pp. 271--282.

\bibitem{Oh2013}
J.~Oh, A.~Zajic, and M.~Prvulovic, ``{Traffic steering between a low-latency
  unswitched TL ring and a high-throughput switched on-chip interconnect},'' in
  \emph{Proceedings of the PACT}, 2013, pp. 309--318.

\bibitem{AbadalONDM}
S.~Abadal, A.~Cabellos-aparicio, J.~A. L{\'{a}}zaro, M.~Nemirovsky,
  E.~Alarc{\'{o}}n, and J.~Sol{\'{e}}-Pareta, ``{Area and Laser Power
  Scalability Analysis in Photonic Networks-on-Chip},'' in \emph{17th
  International Conference in Optical Network Design and Modeling (ONDM)},
  2013.

\end{thebibliography}

\begin{IEEEbiography}[{\includegraphics[width=1in,height=1.25in,clip,keepaspectratio]{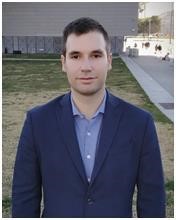}}]{Xavier Timoneda} is a research assistant at the the Universitat Politècnica de Catalunya, where he obtained his degree in Telecommunications Systems Engineering in 2018, after performing his final Degree Thesis at the University of Illinois at Urbana-Champaign (UIUC). He has
authored 4 scientific publications during his first year as a research assistant, and has recently coauthored a book chapter. His research interests include artificial intelligence and chip-scale communications, currently focusing his research in the Development of Artificial Neural Networks for the assembly of Software-driven Functional Metasurfaces.
\end{IEEEbiography}

\begin{IEEEbiography}[{\includegraphics[width=1in,height=1.25in,clip,keepaspectratio]{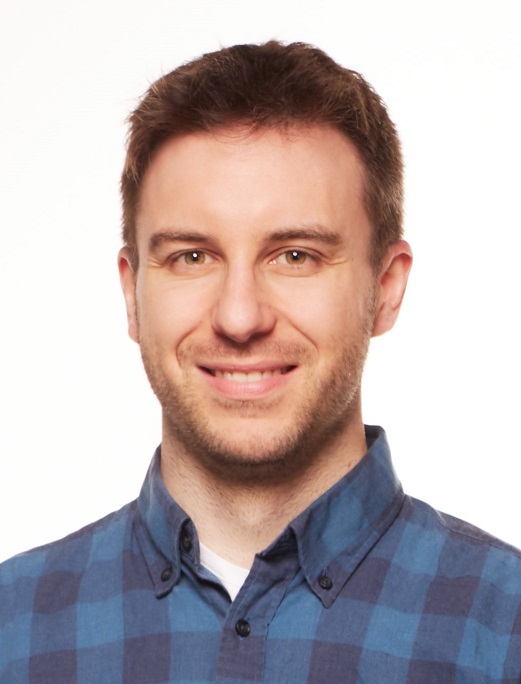}}]{Sergi Abadal} (M'16) received the B.Sc. and M.Sc. degrees in telecommunication engineering from the Universitat Polit\`{e}cnica de Catalunya (UPC), Barcelona, Spain, in 2010 and 2011, and the Ph.D. in computer architecture from the same institution in 2016. During his Ph.D., he was awarded by INTEL within its Doctoral Student Honor Program. He currently works as a postdoctoral researcher at the NaNonetworking Center in Catalunya (N3Cat) and, since October 2019, as Principal Investigator of the EU H2020 project WiPLASH. From 2009 to 2010, he was a Visiting Researcher with the Broadband Wireless Networking Laboratory, Georgia Institute of Technology, Atlanta, USA. He has also been visiting researcher at the School of Computer Science, University of Illinois, Urbana-Champaign, in 2015 and 2018. He has co-authored more than 60 research papers and 7 book chapters. Since 2018, he is Associate Editor of the Nano Communication Networks (Elsevier) Journal, where he was appointed Editor of the Year in 2019. His current research interests are ultra-high-speed on-chip wireless networks and broadcast-enabled manycore processor architectures.
\end{IEEEbiography}

\begin{IEEEbiography}[{\includegraphics[width=1in,height=1.25in,clip,keepaspectratio]{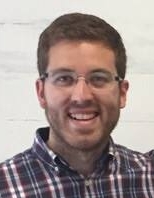}}]{Antonio Franques} is a PhD student in Computer Science at the University of Illinois at Urbana-Champaign (UIUC), and a member of the i-acoma group. His research focuses on the application of high-frequency wireless on-chip communication in manycore architectures. Specifically, his goal is to design new shared-memory architectures that reduce the large cost of core-to-core communication in parallel computing. While working towards his PhD, he also interned twice at AMD Research, working on prototype communication hardware for exascale computing. Prior to joining UIUC, he obtained a Bachelor’s Degree in Telecommunications Engineering from the Polytechnic University of Valencia (UPV), Spain, and also performed two years of research in the area of Computational Mathematics under the supervision of Professors Juan Ramon Torregrosa and Alicia Cordero, as a member of the DAMRES group.
\end{IEEEbiography}

\begin{IEEEbiography}[{\includegraphics[width=1in,height=1.25in,clip,keepaspectratio]{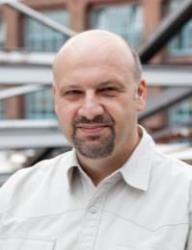}}]{Dionysios Manessis} possesses M.Sc. and Ph.D degrees in Materials Science \& Engineering from Stevens Institute of Technology, NJ, USA and project leadership certificate degrees from Cornell University, NY, USA. He has worked as Technologist for Universal Instruments Corporation in NY, USA and since 2001 has been Senior Technology Scientist in Fraunhofer IZM in Berlin. His main research interests lie on Fine pitch Flip chip and Wafer Level CSP bumping, solder balling, materials selection for advanced packaging technologies, embedding processes for heterogeneous integration of components in PCBs and optical PCBs, large scale prototype manufacturing. In the above technical fields, he has published extensively in international conferences and peer-reviewed journals.
\end{IEEEbiography}

\begin{IEEEbiography}[{\includegraphics[width=1in,height=1.25in,clip,keepaspectratio]{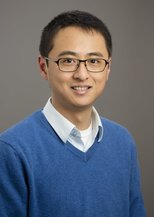}}]{Yin Zhou} is an Assistant Professor at the Electrical and Computer Engineering department of the University of Illinois at Urbana-Champaign (UIUC). He received a B.S. degree in electronics science and technology from Wuhan University, Wuhan, China, in 2008, a M.S. degree in microelectronics from Fudan University, Shanghai, China, in 2011, and a Ph.D. degree in electrical engineering from Columbia University, New York, NY, USA, in 2017. From 2011 to 2012, he also worked as an RF integrated circuits design engineer with MediaTek Singapore. Dr. Zhou is a recipient of the 2015-2016 Qualcomm Innovation Fellowship and the 2015-2016 IEEE Solid-State Circuits Society Predoctoral Achievement Award. He received the Eli Jury Award from the Department of Electrical Engineering at Columbia University in 2016 for his outstanding achievement in the areas of systems, communications, signal processing, or circuits.
\end{IEEEbiography}

\begin{IEEEbiography}[{\includegraphics[width=1in,height=1.25in,clip,keepaspectratio]{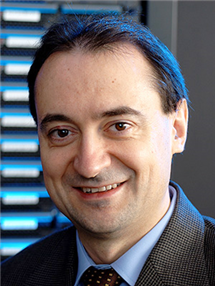}}]{Josep Torrellas} is the Saburo Muroga Professor of Computer Science at the University of Illinois at Urbana-Champaign (UIUC). He is the Director of the Center for Programmable Extreme Scale Computing, and past Director of the Illinois-Intel Parallelism Center (I2PC). He is a Fellow of IEEE (2004), ACM (2010), and AAAS (2016). He received the IEEE Computer Society 2015 Technical Achievement Award, for ``Pioneering contributions to shared-memory multiprocessor architectures and thread-level speculation", and the 2017 UIUC Campus Award for Excellence in Graduate Student Mentoring. He is a member of the Computing Research Association (CRA) Board of Directors. He has served as the Chair of the IEEE Technical Committee on Computer Architecture (TCCA) (2005-2010) and as a Council Member of CRA's Computing Community Consortium (CCC) (2011-2014). He was a Willett Faculty Scholar at UIUC (2002-2009). As of 2016, he has graduated 36 Ph.D. students, who are now leaders in academia and industry. He received a Ph.D. from Stanford University.
\end{IEEEbiography}

\begin{IEEEbiography}[{\includegraphics[width=1in,height=1.25in,clip,keepaspectratio]{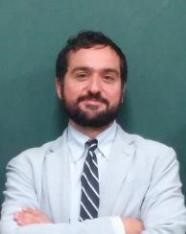}}]{Eduard Alarc\'on} is an associate professor at the the Universitat Politècnica de Catalunya, where he obtained his PhD in electrical engineering in 2000. He has coauthored more than 400 scientific publications, 8 book chapters and 12 patents. He was elected IEEE CAS society distinguished lecturer, member of the IEEE CAS Board of Governors (2010-2013), Associate Editor for IEEE TCAS-I, TCAS-II, JOLPE, and Editor-in-Chief of JETCAS. His research interests include nanocommunications and wireless energy transfer.
\end{IEEEbiography}

\begin{IEEEbiography}[{\includegraphics[width=1in,height=1.25in,clip,keepaspectratio]{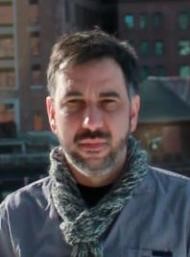}}]{Albert Cabellos} is an assistant professor at Universitat Politècnica de Catalunya, where he obtained his PhD in computer science engineering in 2008. He is co-founder and scientific director of the NaNoNetworking Center in Catalunya. He has been a visiting researcher at Cisco Systems and Agilent Technologies and a visiting professor at the KTH, Sweden, and the MIT, USA. He has co-authored more than 80 research papers. His research interests include nanocommunications and software-defined networking.
\end{IEEEbiography}


\end{document}